\documentclass[11pt]{amsart}

\usepackage{amssymb}
\usepackage{amsmath}
\usepackage{amsbsy}
\usepackage{amsthm}
\usepackage{amscd}
\usepackage{amsfonts}
\usepackage{graphicx}
\usepackage{subfigure}
\usepackage{color}
\usepackage{comment}

\usepackage{epsfig,amsfonts}
\usepackage{graphicx,epstopdf}
\usepackage{psfrag}
\usepackage{array}
\usepackage{color,soul}
\soulregister\cite{7}
\soulregister\citep{7}
\soulregister\ref{7}
\soulregister\pageref{7}
\usepackage[usenames,dvipsnames]{xcolor}
\sethlcolor{yellow}
\usepackage{appendix}

\usepackage{enumerate}
\usepackage{lineno}
\usepackage{placeins}
\usepackage{booktabs,ctable,multirow,longtable}
\usepackage{changepage}

\usepackage[linesnumbered,ruled]{algorithm2e}

\SetCommentSty{mycommfont}
\usepackage{setspace}

\DeclareMathAlphabet{\mathpzc}{OT1}{pzc}{m}{it}
\DeclareMathAlphabet{\mathcalligra}{OT1}{calligra}{m}{it}
\newcommand{\eqn}[2]{\begin{equation} \label{#1} {#2} \end{equation}}

\newcommand{\bs}[1]{\boldsymbol{#1}}

\theoremstyle{definition}

\theoremstyle{remark}

\graphicspath {
    {./}
    {./Figures/}
    {./Images/}
    {./Images/EPS/NIU/}
}

\begin{document}

\title[Data-Driven Fracture Mechanics]{Data-Driven Fracture Mechanics}

\author{P.~Carrara$^{*}$}
\address[P.~Carrara$^{*}$]{Dept. of Mechanical and Process Engineering, ETH Z\"urich, Tannenstr. 3, 8092 Zurich, Switzerland }
\email{pcarrara@ethz.ch}
\thanks{$^{*}$\textit{Corresponding author} pcarrara@ethz.ch}

\author{L.~De Lorenzis}
\address[L.~De Lorenzis]{Dept. of Mechanical and Process Engineering, ETH Z\"urich, Tannenstr. 3, 8092 Zurich, Switzerland }
\email{ldelorenzis@ethz.ch}

\author{L.~Stainier}
\address[L.~Stainier]{Institut de Recherche en G\'enie Civil et M\'ecanique (GeM - UMR 6183), \'Ecole Centrale de Nantes, 1 rue de la No\"e - BP 92101, 44321 Nantes cedex 3, France.}
\email{laurent.stainier@ec-nantes.fr}

\author{M.~Ortiz}
\address[M.~Ortiz]{Division of Engineering and Applied Science, California Institute of Technology, 1200 East California Boulevard, Pasadena, CA 91125, USA.}
\email{ortiz@caltech.edu}


\begin{abstract}
We present a new data-driven paradigm for variational brittle fracture mechanics. The fracture-related material modeling assumptions are removed and the governing equations stemming from variational principles are combined with a set of discrete data points, leading to a model-free data-driven method of solution. The solution at a given load step is identified as the point within the data set that best satisfies either the Kuhn-Tucker conditions stemming from the variational fracture problem or global minimization of a suitable energy functional, leading to data-driven counterparts of both the local and the global minimization approaches of variational fracture mechanics. Both formulations are tested on different test configurations with and without noise and for Griffith and R-curve type fracture behavior.
\end{abstract}

\maketitle

 \tableofcontents

\section{Introduction}
Data-driven techniques rooted in data science and machine learning recently experienced a tremendous development and a boost of applications in many fields such as finance, advertising and marketing, to create predictive models based on large sets of discrete data \cite{Bessa2017}. Related approaches were applied since the late '80s to mechanics problems but were restricted to pre- or post-processing procedures, with the aim of identifying unknown parameters in pre-defined material constitutive laws or for design optimization \cite{Goldberg1987, Ghaboussi1991, Pernot1999}, whereas the solution of the mechanics boundary value problems followed the conventional lines.

More recently, the novel paradigm of model-free data-driven computational mechanics was advocated \cite{Kirchdoerfer2016}. The main idea is that boundary value problems in mechanics are based upon two types of relationships: an epistemic and certain set of basic conservation laws (e.g., energy balance, equilibrium, compatibility) and an empirical and uncertain set of material constitutive equations \cite{Kirchdoerfer2016}. The uncertainty of the latter stems from the attempt to distillate analytical models from collected data, with an unavoidable manipulation of the information \cite{Lopez2018}. In this process, uncertain assumptions on the characteristics of the constitutive model are introduced to obtain objective functions that are calibrated using the collected data, also affected by uncertainty. The consequent  interaction between these two sources of uncertainty is hardly predictable and can be avoided by directly replacing the classical constitutive relationships with information supplied by discrete raw observations.

The data-driven solver in \cite{Kirchdoerfer2016} assigns to each material state the point in the available data set closest to the subset of points fulfilling compatibility and equilibrium. Subsequent extensions were proposed to geometrically nonlinear elasticity \cite{Nguyen2018,Conti2020}, and elastodynamics \cite{Kirchdoerfer2018}. Further developments include a maximum entropy scheme increasing robustness with respect to outliers \cite{Kirchdoerfer2017}, and the reformulation of the problem in the framework of mixed-integer quadratic optimization \cite{Kanno2019}. Based on the approach in \cite{Kirchdoerfer2016}, a new methodology to identify material parameters and stresses in experimental testing based on digital image correlation was developed in \cite{Leygue2018,Stainier2019}. Alternative data-driven formulations \cite{Ibanez2017,Ibanez2017a} seek to reconstruct a constitutive manifold from data using manifold learning methods. In the case of elasticity, the goal is to use data to identify a suitable approximation of the strain energy density functional.

Extensions of the data-driven formulation to inelastic materials have been considered by Eggersmann {\sl et al.} \cite{Eggersmann2018}. The fundamental challenge is to account for the history dependence of the material without modeling assumptions such as an {\sl ad hoc} choice of internal variables. Eggersmann {\sl et al.} \cite{Eggersmann2018} investigate three representational paradigms for the evolving material data sets: i) materials with memory, i.~e., conditioning the material data set to the past history of deformation; ii) differential materials, i.~e., conditioning the material data set to short histories of stress and strain; and iii) history variables, i.~e., optimally identifying variables encoding as much information as possible about history dependence in stress-strain data. In this latter vein, a particular choice of internal variable, namely, the plastic-strain rate, has been considered in \cite{Ladeveze2019}. Despite these advances, the extension of the model-free data-driven approach to dissipative inelastic behavior remains a largely open and non-trivial challenge.

In this paper, we propose a data-driven approach to the solution of the rate independent fracture problem in brittle materials. This class of problems is particularly suited to be adapted to the model-free data-driven paradigm since the natural choice for the history variable is the crack extension, which is easy to measure experimentally. We assume a known linear elastic constitutive behavior of the material and focus on the data-driven solution of the fracture problem. Along the same lines as in \cite{Kirchdoerfer2016}, we remove the fracture-related material modeling assumptions and let the fracture constitutive behavior be fully encoded in a discrete set of material data. In addition, we derive the epistemic and certain set of conservation laws from variational principles. In the variational formulation of the fracture problem, we consider both the stationarity condition of the free energy, i.~e., a solution based on metastability or local stability \cite{Negri2008}, which is the closest to Griffith's view of fracture, and a solution based on global stability in the spirit of \cite{Francfort2008}. The solution at a given load step is identified as the point within the data set that best fulfills either stationarity or global minimization of the free energy, leading to the data-driven counterparts of both the local and the global minimization approaches.

While this paper focuses on the simplest possible setting to best clarify the concepts, several extensions are possible and will be addressed in future research. They include the detailed discussion of different crack modes and the prediction of crack paths in the general two- and three-dimensional case, as well as rate-dependent and ductile fracture and many other more complex cases.

The remainder of this paper is structured as follows. In Sect.~\ref{sct:formulation} standard rate-independent fracture mechanics is briefly recalled along with its variational setting in terms of both local and global minimization of the free energy. The data-driven counterparts are formulated in Sect.~\ref{sct:DD_frac}, where also the main aspects of the numerical implementation are described. Sect.~\ref{sct:NUM_ex} presents some numerical examples, where standard and data-driven formulations are tested on different setups with and without noise, for Griffith and R-curve type fracture behavior. Conclusions are drawn in Sect.~\ref{sct:conclusions}.

\section{Classical fracture mechanics}\label{sct:formulation}

Throughout this work, we are concerned with the problem of modeling the growth of cracks in elastic materials. In order to focus attention on how data-driven concepts apply to fracture mechanics, we consider the simplest possible case of a solid with a known linearly elastic constitutive behavior, in which the crack set can be characterized by a single length parameter $a$.

\subsection{\label{subsec:cracked_solid}Equilibrium of a cracked elastic solid}

\begin{figure}[!hb]
	\begin{adjustwidth}{-3cm}{-3cm}
	\centering
		\includegraphics{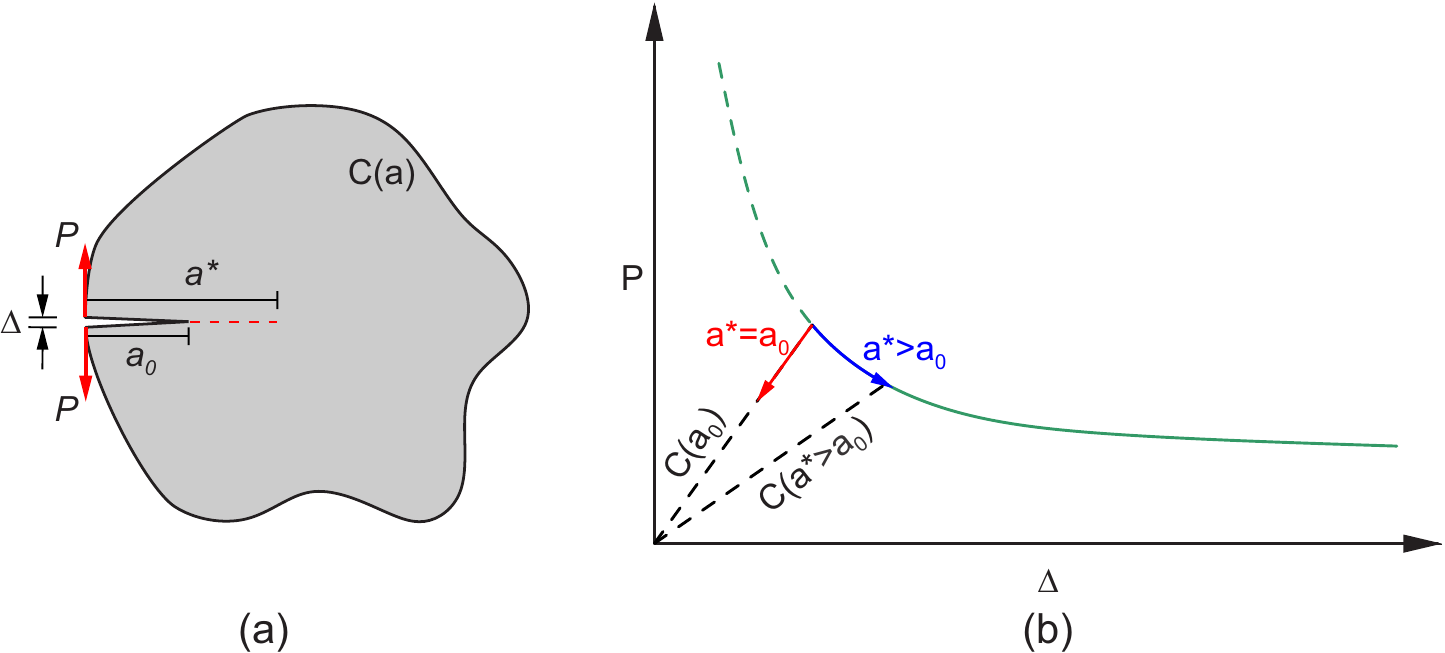}
	\end{adjustwidth}
		\caption{(a) Sketch of the equilibrium problem and (b) schematic representation of the equilibrium curve exhibiting loading/elastic unloading behavior.}
		\label{fig:scheme}
	\end{figure}

Crack growth is an inelastic process in which the crack configuration
evolves according to kinetics and is driven by energetic forces. In
order to exhibit the structure of the theory in its simplest form,
we adopt a compliance representation of the energy and assume planar
rate-independent crack growth under mode I loading. Under these conditions,
the configuration of the crack is described by the single variable
$a$, which plays the role of a state variable, the loading by
an effective force $P$ and the deformation by a conjugate effective
displacement $\Delta$. At equilibrium $P$ and $\Delta$ necessarily
bear a linear relation of the form
\begin{equation}
\Delta=C(a)P\label{compl}
\end{equation}
where $C(a)$ is the crack-length-dependent \textsl{compliance} of
the solid (Fig.~\ref{fig:scheme}a).  Throughout this paper we assume the compliance function to be exactly known.

We assume displacement control and formulate the equilibrium problem
as follows. Suppose the solid has an initial crack of length $a_{0}$,
and it is imparted a prescribed opening displacement $\Delta$ (Fig.~\ref{fig:scheme}a). Our
goal is to determine the equilibrium crack length $a^{*}\left(\Delta,a_{0}\right)$,
where we postulate $a^{*}\geq a_{0}$ due to irreversibility, and
the corresponding load $P$ (Fig.~\ref{fig:scheme}a). More generally, within an incremental
displacement-controlled loading process, we may want to determine
the incremental behavior of the system; i.e., for given $\Delta_{k+1}$
at time $t_{k+1}$ and $a_{k}$ at time $t_{k}$, determine $a_{k+1}\left(\Delta_{k+1},a_{k}\right)$,
with $a_{k+1}\geq a_{k}$ due to irreversibility, and
the corresponding load $P_{k+1}$. Note that the quantity $a_k$ plays here the role of a history variable.

Both the equilibrium and the incremental problem can be characterized
variationally by means of energy dissipation principles. As
follows, we formulate the equilibrium problem. We introduce the elastic
strain energy of the solid

\begin{equation}\label{elstr}
E=\frac{\Delta^{2}}{2C(a)}\,,
\end{equation}
which has the properties
\begin{equation}\label{P_G}
P=\frac{\partial E}{\partial\Delta}(\Delta,a),\qquad G=-\frac{\partial E}{\partial a}(\Delta,a)\,,
\end{equation}
where $G$ is known as the energy release rate. Combining (\ref{elstr})
and (\ref{P_G}b) gives

\begin{equation}
G=\frac{\Delta^{2}}{2C^{2}(a)}\frac{dC}{da}\,,
\end{equation}
which provides a shortcut to compute $G$ known as compliance method.
Let us now introduce the free energy
\begin{equation}
F(\Delta,a)=E(\Delta,a)+F_{R}(a)\,,
\end{equation}
where the resistance term $F_{R}(a)$ has the property that
\begin{equation}
G_{R}(a)=\frac{dF_{R}}{da}(a)\,,\label{eq:G_R}
\end{equation}
defines the resistance curve in terms of energy release rates. For
the solution of the problem based on the variational principle of
free energy minimization we have two options.

One possibility is to look for the value of crack length which corresponds
to the \textit{global minimum} of the free energy, i.~e.,
\begin{equation}
F\left(\Delta,a^{*}\right)\leq F\left(\Delta,a\right)\quad\forall a\geq a_{0}\,,
\end{equation}
or

\begin{equation}\label{eq:glob}
a^{*}\left(\Delta,a_{0}\right)=\mathrm{argmin\left\{ \mathit{F\left(\Delta,a\right):a\geq a_{0}}\right\} }\,.
\end{equation}

Another possibility is to look for the value of crack length which
corresponds to a \textit{local minimum} of the free energy, i.~e., to look for $a^{*}$ such that there exists $h>0$ satisfying

\begin{equation}
F\left(\Delta,a^{*}\right)\leq F\left(\Delta,a^{*}+a-a^{*}\right)\qquad\forall a\geq a_{0},\left|a-a^{*}\right|\leq h\,,
\end{equation}
i.~e., for each $a\geq a_{0}$ in the \emph{neighborhood} of $a^{*}$. A first-order Taylor series expansion of the right-hand side, along
with (\ref{P_G}b) and (\ref{eq:G_R}), leads to
\begin{equation}\label{eq:ineq}
\left[G\left(\Delta,a^{*}\right)-G_{R}\left(a^{*}\right)\right]\left(a-a^{*}\right)\leq0\quad\forall a\geq a_{0}\,.
\end{equation}
There are two possibilities to satisfy this condition, namely
\begin{itemize}
\item[-] for $a^{*}>a_{0}$, $a-a^{*}$ can have any sign. Hence, in order to satisfy
(\ref{eq:ineq}), it must be $G\left(\Delta,a^{*}\right)-G_{R}\left(a^{*}\right)=0$;
\item[-] for $a^{*}=a_{0}$, it can only be $a-a^{*}\geq0$, so that (\ref{eq:ineq})
gives $G\left(\Delta,a^{*}\right)-G_{R}\left(a^{*}\right)\leq0$.
\end{itemize}
These conditions can be collected in Kuhn-Tucker form as

\begin{eqnarray}\label{eq:KT}
a^{*}-a_{0} & \geq & 0\nonumber \\
G\left(\Delta,a^{*}\right)-G_{R}\left(a^{*}\right) & \leq & 0\\
\left[G\left(\Delta,a^{*}\right)-G_{R}\left(a^{*}\right)\right]\left(a^{*}-a_{0}\right) & = & 0\,.\nonumber
\end{eqnarray}
Evidently, these relations reduce to the Griffith criterion when $G_{R}(a)=G_{c}=$
constant.

Global and local minimization obviously deliver the same solution
if the free energy is a convex function of $a$. Otherwise, it is
known in the literature that the two options in general deliver different
results. In particular, the  local minimality principle forbids the evolution between states that are separated by energetic barriers and tries to find the next equilibrium state in the neighborhood of the current state. Conversely, globally stable processes can evolve to conditions that are energetically more favorable but separated by arbitrarily large energetic barriers violating the causality principle and leading to anticipated jumps. This ultimately leads to a globally stable states domain that is smaller compared to the local counterpart, see \cite{Negri2008}.

Inserting $a^{*}\left(\Delta,a_{0}\right)$ obtained either from (\ref{eq:glob})
or from (\ref{eq:KT}) into the equilibrium relation (\ref{P_G}a),
we obtain

\begin{equation}\label{eq:P_eq}
P=\frac{\partial E}{\partial\Delta}(\Delta,a^{*}\left(\Delta,a_{0}\right))\,,
\end{equation}
which characterizes all possible $\left(P,\Delta\right)$ pairs attainable
by the system given $a_{0}$. Hence, for a given $\Delta$ it is immediate
to find $P$. We expect the curve defined by (\ref{eq:P_eq}) to exhibit loading-unloading
behavior, with elastic unloading occurring when $a^{*}(\Delta,a_{0})=a_{0}$,
i.~e., under conditions of crack arrest, and loading when $a^{*}(\Delta,a_{0})>a_{0}$,
i.~e., under conditions of crack growth (Fig.~\ref{fig:scheme}b).

\subsection{\label{subsec:cracksol_test}Equilibrium of a cracked elastic solid
connected to a testing machine}

Suppose, more generally, that the solid is connected to a testing
machine of compliance $C_{{\rm M}}$. Then, the opening displacement
$\Delta$ and the load are related as
\begin{equation}
\Delta_{{\rm T}}=\Delta+C_{{\rm M}}P,\label{machine}
\end{equation}
where $\Delta_{{\rm T}}$ is a time-dependent control opening displacement.
Other loading devices result in different load-displacement relations
and may involve different control parameters, but the simple linear
device (\ref{machine}) suffices to illustrate the concept. The elimination
of $P$ between (\ref{P_G}a) and (\ref{machine}) delivers

\begin{equation}\label{eq:delta}
\Delta=\frac{C\left(a\right)}{C\left(a\right)+C_{M}}\Delta_{T}\,.
\end{equation}

Once again we formulate the equilibrium problem. The solid has an
initial crack of length $a_{0}$, and it is subjected by the testing
machine to a prescribed opening displacement $\Delta_{T}$. Our goal
is to determine the equilibrium crack length $a^{*}\left(\Delta_{T},a_{0}\right)$,
where we postulate $a^{*}\geq a_{0}$ due to irreversibility, and
the corresponding load $P$ and opening displacement $\Delta$. In
fact, we first determine $a^{*}\left(\Delta,a_{0}\right)$ to leverage
the results in Sect. \ref{subsec:cracked_solid}, and then show
how one can directly determine $a^{*}\left(\Delta_{T},a_{0}\right)$.

In this case one can introduce a mixed variational principle, as follows.
Define the function
\begin{equation}
\Phi(\Delta,P,a)=E(\Delta,a)+F_{R}(a)-\frac{C_{{\rm M}}}{2}P^{2}-P(\Delta-\Delta_{{\rm T}}).\label{eq:mixed_varpr}
\end{equation}
It is straightforward to verify that the Euler-Lagrange equations
of the saddle-point problem
\begin{equation}\label{minimax}
\min_{\Delta}\min_{a\geq a_{0}}\max_{P}\Phi(\Delta,P,a)\,,
\end{equation}
are indeed (\ref{eq:P_eq}), (\ref{machine}) and the Kuhn-Tucker
conditions (\ref{eq:KT}). The latter conditions obviously only hold
if local minimization is pursued; alternatively, the direct global
minimization of (\ref{eq:mixed_varpr}) with respect to $a\geq a_{0}$
must be carried out. Evidently, once, for a given $\Delta_{T}$, $a^{*}\left(\Delta,a_{0}\right)$
is determined by either global or local minimization, the new state
$(P,\Delta)$ of the system must satisfy relations (\ref{eq:P_eq})
and (\ref{machine}) simultaneously, i.~e., it must lie at the intersection
of the curves defined by (\ref{eq:P_eq}) and (\ref{machine}). This defines automatically
$P$ and $\Delta$.

A more direct way to proceed is the following. By substituting (\ref{P_G}a)
and (\ref{machine}) into (\ref{eq:mixed_varpr}), the following reduced
function is obtained

\begin{equation}\label{eq:mixed_varpr-1}
\tilde{\Phi}(\Delta_{T},a)=\tilde{E}(\Delta_{T},a)+F_{R}(a)\,,
\end{equation}
with
\begin{equation}\label{elstr-1}
\tilde{E}=\frac{1}{2}\frac{\Delta_{T}^{2}}{C(a)+C_{M}}\,,
\end{equation}
which can then be directly minimized (globally or locally) with respect
to $a$ at given $\Delta_{T}$ and $a_{0}$. Global minimization delivers

\begin{equation}
a^{*}\left(\Delta_{T},a_{0}\right)=\mathrm{argmin\left\{ \mathit{\tilde{\Phi}\left(\Delta_{T},a\right):a\geq a_{0}}\right\} }\,,\label{eq:glob-1}
\end{equation}
whereas local minimization leads to

\begin{eqnarray}
a^{*}-a_{0} & \geq & 0\nonumber \\
\tilde{G}\left(\Delta_{T},a^{*}\right)-G_{R}\left(a^{*}\right) & \leq & 0\label{eq:KT-1}\\
\left[\tilde{G}\left(\Delta_{T},a^{*}\right)-G_{R}\left(a^{*}\right)\right]\left(a^{*}-a_{0}\right) & = & 0\,,\nonumber
\end{eqnarray}
where $\tilde{G}\left(\Delta_{T},a\right)$ is obtained from the combination
of $G\left(\Delta,a\right)$ and (\ref{eq:delta}).

\section{Data-driven fracture mechanics} \label{sct:DD_frac}

For most materials, the crack resistance curve is only known through
material testing, which results in limited data in the form of point
sets. In addition, such data can be noisy. In the spirit of data-driven
mechanics \cite{Kirchdoerfer2016}, we explore the possibility of solving fracture
mechanics problems directly from data.

\subsection{Data representation} \label{sct:full}

The question now arises of how to characterize fracture by means of
data. To this end, we begin by reexamining the classical equilibrium
problem of Sect. \ref{subsec:cracksol_test} in which $E(\Delta,a)$,
$F_{R}(a)$ and the loading device are exactly known and characterized.

We adopt the viewpoint that a solution is a pair $(\Delta,P)$ that
simultaneously satisfies compatibility, equilibrium and the material
laws. The space of solutions, or \textsl{phase space}, is therefore
the space $\mathcal{Z}$ of work-conjugate pairs $(\Delta,P)$. Given
$a_{0}$, we may identify the subset of $\mathcal{Z}$,
\begin{equation}
\mathcal{D}=\{(\Delta,P)\,:\,\text{equation (\ref{eq:P_eq})}\},
\end{equation}
as a \textsl{material data set} which collects the net sum of our
knowledge about the material. Likewise, given $\Delta_{{\rm T}}$
we may identify the subset of $\mathcal{Z}$,
\begin{equation}
\mathcal{E}=\{(\Delta,P)\,:\,\text{equation (\ref{machine})}\},
\end{equation}
as a \textsl{constraint set}, consisting of all points in phase space
that are compatible and in equilibrium with the loading machine. Evidently,
within this framework the solution set for given $a_{0}$ and $\Delta_{{\rm T}}$
is the intersection $\mathcal{D}\cap\mathcal{E}$, i.~e., the collection
of points in phase space that are simultaneously in the material and
constraint data sets.

Suppose now that the elastic energy $E(\Delta,a)$ is exactly known,
e.~g., in terms of a fully characterized compliance, whereas the
resistance part $F_{R}(a)$ of the free energy or its derivative $G_{R}(a)$
is known only from data. Given a crack length $a_{0}$ and an opening
displacement $\Delta$, the corresponding equilibrium crack length
$a^{*}(\Delta,a_{0})$ can be found from the data by either global or local minimization.
Within an incremental loading procedure, we compute the equilibrium crack length $a_{k+1}$
for given displacement $\Delta_{k+1}$ at time $t_{k+1}$ and crack
length $a_{k}$ at time $t_{k}$. This relation in turn defines the
material data set,
\begin{equation}
\mathcal{D}_{k+1}=\left\{(\Delta_{k+1},P(\Delta_{k+1},a^{*}(\Delta_{k+1},a_{k}))\right\}\,,
\end{equation}
of all possible opening displacements and loads attainable at time
$t_{k+1}$.

Let $\mathcal{E}_{k+1}$ be the constraint set of points $(\Delta_{k+1},P_{k+1})$
consistent with the loading device at time $t_{k+1}$. We define the
data-driven solution set at time $t_{k+1}$ as the intersection $\mathcal{D}_{k+1}\cap\mathcal{E}_{k+1}$,
i.~e., the collection of points in phase space that are simultaneously
in the material and constraint data sets.

As in (\ref{minimax}), the entire solution $(\Delta_{k+1},P_{k+1},a_{k+1})$
can be characterized jointly by means of the saddle point problem
\begin{equation}
\min_{\Delta}\min_{{a\geq a_{0}\atop (a,F_{R})\in\mathcal{D}_{R}}}\max_{P}\Big(E(\Delta,a)+F_{R}-\frac{C_{{\rm M}}}{2}P^{2}-P(\Delta-\Delta_{{\rm T}})\Big),\label{eq:minimax_data}
\end{equation}
which now entails a discrete minimization over the resistance data
set $\mathcal{D}_{R}$. If using local minimization with respect to
$a$, $(a,F_{R})\in\mathcal{D}_{R}$ must be substituted by $(a,G_{R})\in\mathcal{D}_{R}$
in (\ref{eq:minimax_data}).

\subsection{Computational procedure} \label{sct:direct}

A more direct and practical way of proceeding, which mirrors the introduction of
the ``reduced'' function $\tilde{\Phi}$ in Sect. \ref{subsec:cracksol_test},
is the following.

When using the global minimization approach, the resistance data set
$\mathcal{D}_{R}$ consists of pairs $(\hat{a}_i,\hat{F}_{Ri})$ and
minimization is straightforwardly performed as follows: $a^{*}(\Delta_{T},a_{0})=\hat{a}_{i^{*}}$ with
\begin{equation}
i^{*}(\Delta_{T},a_{0})={\rm \underset{\mathit{i}}{argmin}}\left\{\tilde{E}(\Delta_{T},\hat{a}_{i})+\hat{F}_{Ri}\,:\,\hat{a}_{i}\geq a_{0},\ (\hat{a}_{i},\hat{F}_{Ri})\in\mathcal{D}_{R}\right\}.\label{data_glob-1}
\end{equation}
Note that $a^{*}(\Delta_{T},a_{0})$ follows from a discrete
minimization over the point set $\mathcal{D}_{R}$ and is, therefore,
stepwise.

With local minimization, the data set $\mathcal{D}_{R}$ consists of
pairs $(\hat{a}_i,\hat{G}_{Ri})$. Finding $a^{*}(\Delta_{T},a_{0})$ requires a little more care. A possible strategy is based on closest-point projection. Here we first determine $G_{R0}$ as the value of $\hat G_R$ corresponding to $\hat a = a_0$. Then, if the propagation condition is met, i.~e., $G_{R0}-\tilde{G}\left(\Delta_{T},a_0\right)\le0$, and for
each $(\hat{a}_i,\hat{G}_{Ri})$ pair in the data set with $\hat{a}_{i}\geq a_{0}$, we determine
the distance to the analytical curve $\tilde{G}\left(\Delta_{T},a\right)$
as

\begin{equation}
\tilde{d}_{i}\left(\Delta_{T},a_{0}\right)={\rm \underset{\mathit{a\geq a_{0}}}{min}}\left\{\sqrt{\left(\hat{a}_{i}-a\right)^{2}+\left(\hat{G}_{Ri}-\tilde{G}\left(\Delta_{T},a\right)\right)^{2}}\right\}\,,
\end{equation}
and then compute $a^{*}(\Delta_{T},a_{0})$ by minimizing the distance
between $\tilde{G}\left(\Delta_{T},a\right)$ and the resistance data set,
as follows: $a^{*}(\Delta_{T},a_{0})=\hat{a}_{i^{*}}$ with

\begin{equation}
i^{*}(\Delta_{T},a_{0})={\rm \underset{\mathit{i}}{argmin}}\left\{\tilde{d}_{i}(\Delta_{T},a_{0})\,:\,\hat{a}_{i}\geq a_{0},\ (\hat{a}_{i},\hat{G}_{Ri})\in\mathcal{D}_{R}\right\}.\label{data_loc-2}
\end{equation}
A second possible strategy for local minimization is based on the best approximation of the third Kuhn-Tucker condition (known as the consistency condition). With this strategy, we compute

\begin{equation}
\begin{split}i^{*}(\Delta_{T},a_{0})=&{\rm \underset{\mathit{i}}{argmin}}\left\{\left|\left(\tilde{G}\left(\Delta_{T},\hat{a}_{i}\right)-\hat{G}_{Ri}\right)\left(\hat{a}_{i}-a_{0}\right)\right|\,:\right.\\ &\left.\,\hat{a}_{i}\geq a_{0},\tilde{G}\left(\Delta_{T},\hat{a}_{i}\right)-\hat{G}_{Ri}\leq0,\ (\hat{a}_{i},\hat{G}_{Ri})\in\mathcal{D}_{R}\right\}.\end{split}\label{data_loc-1-1}
\end{equation}
and once again $a^{*}(\Delta_{T},a_{0})=\hat{a}_{i^{*}}$. Note that in (\ref{data_loc-1-1}) the crack arrest/propagation condition is directly imposed within the minimization procedure. In all cases, once $a^{*}$ is known, $\Delta$ and $P$ can be computed directly:

\begin{equation}
\Delta=\frac{C\left(a^{*}\right)}{C\left(a^{*}\right)+C_{M}}\Delta_{T}\,,\quad P=\frac{\Delta}{C\left(a^{*}\right)}\,.
\end{equation}

Within an incremental loading procedure, we perform either global
or local minimization to compute the equilibrium crack length $a_{k+1}$
for given displacement $\Delta_{T\,k+1}$ at time $t_{k+1}$ and crack
length $a_{k}$ at time $t_{k}$. Finally, we can compute $\Delta_{k+1}$
and $P_{k+1}$ directly. More details are given in the next subsection.

\subsection{Numerical implementation} \label{sct:NUM_algo}

This section provides more details on how the data-driven solution of the fracture problem is implemented within an incremental loading procedure.  The applied displacement is parameterized as $\Delta_{T\,k} = \delta_Tk$ where $k$ is the load step number and $\delta_T$ the load step increment. The aim is then, given the current imposed displacement $\Delta_{T\,k+1}$ and the previous crack length $a_k$, to find within the material set $\mathcal{D}_R$ the state that fulfils the global or local data-driven criteria described in Sect.~\ref{sct:direct}. In this context, the variable $a$ acts as a history variable, that is trivially initialized to the length of the initial crack $a_0$.

Both global and local minimization need to be constrained by the irreversibility condition. A convenient way to enforce this constraint is to \emph{a priori} restrict the minimization procedures to the states for which $a_{k+1} \ge a_k$ is fulfilled. Alternatively, one can adopt a reference system attached to the crack tip and reformulate the minimization problems in terms of crack length increments $\Delta a_{k+1} \ge 0$, a choice which makes the data set independent on the geometry and test setup.

In the following it is assumed that the material data set is sufficient to study the problem at hand, meaning that the points in $\mathcal{D}_R$ cover the propagation of the crack from the initial length $a_0$ to the maximum extension allowed by the geometry and setup under investigation.

	\subsubsection{Global minimization}
Algorithm~\ref{algo:DD_GLOB} presents the procedure to obtain the data-driven solution of the crack propagation problem as the global minimum of the function $\tilde{\Phi}(\Delta_T,\,a)$. Note that, if the material data set contains no point with $\hat{a}=a_0$, at the first load step the minimization procedure always predicts crack propagation, and the extent of this propagation depends on the characteristics of the data set $\mathcal{D}_R$.

The global minimization procedure does not detect the occurrence of unstable crack propagation and keeps delivering a (possibly constant) solution state.

\begin{algorithm}
	\DontPrintSemicolon
	\setstretch{1.4}
	$\bs{Step}:\,\,k+1$

	\KwIn{$\Delta_{T\,{k+1}}$, $a_k$}

	\KwOut{{$a_{k+1},\,\,\Delta_{k+1},\,\,P_{k+1}$}}

	\tcc{BEGINNING OF THE COMPUTATION}

	\bf{Define}: $\mathcal{D}_{R\,\,{k+1}}=\left\{(\hat a_i,\,\hat F_{Ri})\in\mathcal{D}_{R}: \hat a\ge a_k\right\}$
	
	\tcc{Compute the solution}
	\bf{Compute}:
	{$i^{*}(\Delta_{T},a_{0})={\rm \underset{\mathit{i}}{argmin}}\left\{\tilde{E}(\Delta_{T},\hat{a}_{i})+\hat{F}_{Ri}\,:\,\ (\hat{a}_{i},\hat{F}_{Ri})\in\mathcal{D}_{R\,\,{k+1}}\right\}$
	$a_{k+1}=\hat{a}_{i^{*}}$ \\
	$\Delta_{k+1} = \displaystyle\frac{C\left(a_{k+1}\right)}{C\left(a_{k+1}\right)+C_{M}}\Delta_{T\,k+1},\,\, P_{k+1} = \displaystyle\frac{\Delta_{T\,k+1}}{C\left(a_{k+1}\right)+C_{M}}$ }

	\caption{Data-driven fracture mechanics algorithm - global minimization. Given: $\,\, \tilde{E}(\Delta_T,\, a)$,  $\,\,C(a)$, $\,\,\mathcal{D}_R$.}
	\label{algo:DD_GLOB}
\end{algorithm}

	\subsubsection{Local minimization based on closest point projection}\label{sct:DD_CPP}
	
The implementation of the closest point projection strategy is detailed in Algorithm~\ref{algo:DD_DIST}. This algorithm is slightly more complicated but it allows a deeper analysis. The history variables are both $a$ and $G_R$ and this allows to distinguish within the data-driven search procedure between an elastic step and a dissipative step. The related energy release rate quantities are collected in $G_{DD\,k+1}$. This seemingly redundant distinction can be exploited to implement a data-driven search also for the elastic step. This can be done by introducing an elastic material data set that accounts, e.g., for the measuring tolerance of the crack length. Moreover, this distinction reduces the computational time if requiring a data-driven search only when the propagation condition is met.

The quantity $G_R$ triggers the crack propagation, hence, the initial crack increment depends on its initialization. If the initial crack length $a_0$ corresponds to a point of the material data set $\mathcal{D}_R$, then the couple $(a_0,\,G_{R0})$ can be trivially set to the coordinates of that point. Otherwise, one can set $G_{R0}$ as the average value of $\hat G$ of the two points in $\mathcal{D}_R$ with value of $\hat a$ immediately larger and smaller than $a_0$ or set $G_{R0}$ to 0. With the former choice, adopted here, the computation involves an elastic branch from the first load step that proceeds until $G(\Delta_T,\,a_0)\le G_{R0}$, the latter choice instead predicts an initial increment of crack length similarly to what was noted for the global minimization approach.

Algorithm~\ref{algo:DD_DIST} entails the definition of a tolerance ($tol$) to be applied to the distance between the energy release rate function and the material data set. This allows to detect the occurrence of unstable crack propagation. The value of $tol$ must be  related to the characteristics of the data set and we suggest to set it to 5-10 times the average crack size increment in the adopted material data set.

\begin{algorithm}
	\DontPrintSemicolon
	\setstretch{1.4}
	$\bs{Step}:\,\,k+1$

	\KwIn{$\,\,\Delta_{T\,k+1}$, $a_{k}\,, G_{R\,k}$}

	\KwOut{{$a_{k+1},\,G_{R\,\,k+1},\,\,\Delta_{k+1},\,P_{k+1},\,\,G_{DD\,k+1}$}}

	\tcc{BEGINNING OF THE COMPUTATION}


	\bf{Define}: $\mathcal{D}_{R\,\,k+1}=\left\{(\hat a_i,\,\hat G_{R\,i})\in\mathcal{D}_{R}: \hat a_i\ge a_k\right\}$
	
	\tcc{Compute the solution}

	\uIf{$G_{R\,k} > \tilde{G}\left(\Delta_{T\,k+1},a_{k}\right)$}{\vspace{-6.5mm}\tcp*{Elastic step}
	
	\bf{Assign}: $a_{k+1}=a_k$,  $G_{R \, k+1}=G_{R \, k}$, $G_{DD \, k+1}= \tilde{G}\left(\Delta_{T\,k+1},a_{k}\right)$}
	\Else{\vspace{-6.7mm}\tcp*{Dissipative step}
	
	\bf{Compute}:
	
	$\tilde{d}_{i}\left(\Delta_{T\,k+1},a_{k}\right)={\rm \underset{\mathit{a\geq a_{k}}}{min}}\left\{\sqrt{\left(\hat{a}_{i}-a\right)^{2}+\left(\hat{G}_{Ri}-\tilde{G}\left(\Delta_{T\,k+1},a\right)\right)^{2}}\right\}$\;

        \vspace{3mm}$i^{*}(\Delta_{T\,k+1},a_{k})={\rm \underset{\mathit{i}}{argmin}}\left\{\tilde{d}_{i}(\Delta_{T\,k+1},a_{k})\,:(\hat{a}_{i},\hat{G}_{Ri})\in\mathcal{D}_{R\,k+1}\right\}$\;

        \vspace{3mm}\bf{Assign}: $a_{k+1}=\hat{a}_{i^{*}}$, $G_{R \, k+1}=\hat{G}_{R \, i^{*}}$, $G_{DD \, k+1}= \hat{G}_{R \, i^{*}}$\;

       \vspace{3mm}\If{$\tilde{d}_{i^{*}}(\Delta_{T\,k+1},\,a_k)> tol$}{
		Unstable propagation $\rightarrow$ EXIT\tcp*{Specimen failure}}
}
		
		{\bf{Compute}: $\Delta_{k+1} = \displaystyle\frac{C\left(a_{k+1}\right)}{C\left(a_{k+1}\right)+C_{M}}\Delta_{T\,k+1},\,\, P_{k+1} = \displaystyle\frac{\Delta_{T\,k+1}}{C\left(a_{k+1}\right)+C_{M}}$ }
	\caption{Data-driven fracture mechanics algorithm - local minimization with closest point projection. Given: $\tilde{G}(\Delta_T,\, a)$, $\,\,C(a)$, $\,\,\mathcal{D}_R$.}
	\label{algo:DD_DIST}
\end{algorithm}

\subsubsection{Local minimization based on the consistency condition}
The implementation of the local minimization based on the consistency condition is detailed in Algorithm~\ref{algo:DD_CONS}. Here, the solution of the problem given a certain displacement $\Delta_{T\,k+1}$ is defined as the point $(\hat a, \hat G_R)$ that minimizes the violation of the consistency condition (\ref{eq:KT-1}c). Such condition already encodes the propagation criterion and, hence, makes the definition of $G_R$ redundant (although still possible) preventing the distinction between elastic and dissipative steps. In turn, this might lead, when $a_0$ is not represented in $\mathcal{D}_R$, to an increment of the crack size at the first load step, similarly to what mentioned for the closest point projection where $G_R$ is initialized to 0. The restriction of the minimization to the points with $\hat G_R \ge  \tilde{G}(\Delta_T,\,\hat a)$ allows to detect unstable crack propagation without the introduction of any tolerance.

\begin{algorithm}
	\DontPrintSemicolon
	\setstretch{1.4}
	$\bs{Step}:\,\,k+1$

	\KwIn{$\,\,\Delta_{T\,k+1}$, $\,\, a_k$}

	\KwOut{{$a_{k+1},\,G_{R\,k+1},\,\Delta_{k+1},\,P_{k+1}$}}

	\tcc{BEGINNING OF THE COMPUTATION}

	\bf{Define}: $\mathcal{D}_{R\,k+1}=\left\{(\hat a_i,\,\hat G_{R\,i})\in\mathcal{D}_{R}: \hat a_i\ge a_k,\, \hat G_{R\,i} \ge  \tilde{G}(\Delta_{T\,k+1},\,\hat a_i)\right\}$
	
	\tcc{Compute the solution}
	\uIf{$\mathcal{D}_{R\,k+1} \neq \varnothing$}{
	
	\bf{Compute}:
	
	$\begin{array}{ll}i^{*}(\Delta_{T\,k+1},a_{k})=&{\rm \underset{\mathit{i}}{argmin}}\left\{\left|\left(\tilde{G}\left(\Delta_{T\,k+1},\hat{a}_{i}\right)-\hat{G}_{Ri}\right)\left(a_{k}-\hat{a}_{i}\right)\right|\,:\right.\\ &\left.(\hat{a}_{i},\hat{G}_{Ri})\in\mathcal{D}_{R\,k+1}\right\}\end{array}$
	
	$a_{k+1}=\hat{a}_{i^{*}}$, $G_{R \, k+1}=\hat{G}_{R \, i^{*}}$}
	
		\Else{
		
		Unstable propagation $\rightarrow$ EXIT\tcp*{Specimen failure}}
		
		{\bf{Compute}: $\Delta_{k+1} = \displaystyle\frac{C\left(a_{k+1}\right)}{C\left(a_{k+1}\right)+C_{M}}\Delta_{T\,k+1},\,\, P_{k+1} = \displaystyle\frac{\Delta_{T\,k+1}}{C\left(a_{k+1}\right)+C_{M}}$ }

	\caption{Data-driven fracture mechanics algorithm - local minimization with consistency condition. Given: $\,\, \tilde{G}(\Delta_T,\, a)$, $\,\,C(a)$,$\,\,\mathcal{D}_R$.}
	\label{algo:DD_CONS}
\end{algorithm}

\section{Numerical examples} \label{sct:NUM_ex}

\begin{figure}[!h]
	\begin{adjustwidth}{0cm}{0cm}
	\centering
		\includegraphics{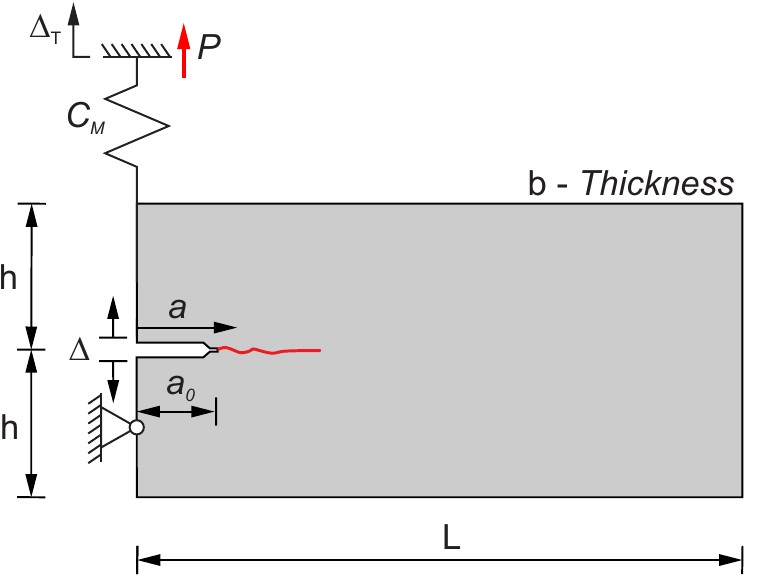}
	\end{adjustwidth}
		\caption{Scheme and geometry of the double cantilever beam test.}
		\label{fig:DCB}
	\end{figure}

To demonstrate the capabilities of the proposed approach let us consider a double cantilever beam (DCB) with dimensions $L\times 2h\times b$ and initial crack length $a_0$, whose arms are subjected to bending (Fig.~\ref{fig:DCB}). The test is performed imposing the loading device displacement $\Delta_T$ (Fig.~\ref{fig:DCB}) following a load ramp $\Delta_{T\,k} = \delta_T\,k$.
If not otherwise specified, the parameters are those in Tab.~\ref{tab:param}.

\begin{table}
\centering
\begin{tabular}{llll}
\toprule
 Young's modulus  & $Y$ & = &  $70$ GPa\\[4pt]
 Height& $h$ & = & $3$ mm   \\[4pt]
 Length & $ L $& = &  $30$ mm  \\[4pt]
 Thickness & $ b $& = &  $1$ mm  \\[4pt]
 Initial crack length & $ a_0 $& = &  $3$ mm  \\[4pt]
Displacement increment & $ \delta_T $& = &  $1.5\cdot10^{-3}$ mm   \\[4pt]
 Machine compliance &  $C_M$& = & $2\cdot 10^{-3}$ mm/N \\ [4pt]
Griffith fracture toughness &  \multirow{2}{*}{$\gamma$}& \multirow{2}{*}{=} & \multirow{2}{*}{$0.06$ N/mm} \\
\hspace{15mm}(Reference)   \\
\toprule
\end{tabular}
\caption{Parameters used for the computations.}
\label{tab:param}
\end{table}

 The following dimensionless quantities are used
	\eqn{eq:dimensionless}{\begin{array}{cccc} \displaystyle\bar L = \frac{L}{L}=1\,, & \displaystyle\bar h = \frac{h}{L}\,, & \displaystyle\bar a = \frac{a}{L}\,, &\displaystyle\bar b = \frac{b}{L}\,, \\[20pt]
	\displaystyle\bar C_M = {C_M\gamma}\,,  & \displaystyle\bar Y = \frac{YL}{\gamma}\,, & \displaystyle\bar \Delta_T = \frac{\Delta_T}{L}\,, & \displaystyle\bar \Phi= \frac{\Phi}{\gamma L^2} \,, \\ [20pt]%
						\displaystyle\bar G= \frac{G}{\gamma}\,, & \displaystyle\bar E= \frac{E}{\gamma L^2}\,, & \displaystyle\bar G_R = \frac{G_R}{\gamma}\,, & \displaystyle\bar F_R= \frac{F_R}{\gamma L^2}\,.					
							\end{array}}

\noindent where the Griffith critical energy release rate $\gamma$ should be intended as a reference value. For the geometry reported in Fig.~\ref{fig:DCB}, the dimensionless energy release rate is

	\eqn{eq:ERR_DCB_nondim}{\bar G(\bar \Delta_T,\,\bar a) = 12 \bar a^2 \bar Y \bar h^3\left[\frac{\bar \Delta_T}{8\bar a^3+\bar C_M \bar Y \bar b \bar h^3}\right]^2\,,}

\noindent while the dimensionless compliance $\bar C(\bar a)$ and applied load $\bar P(\bar\Delta_T,\bar a)$ are

\eqn{eq:dimless_param}{ \bar C (\bar a)=\frac{8\bar a^3}{\bar Y \bar b \bar h^3}\,, \quad\quad \bar P(\bar\Delta_T,\bar a)= \frac{\bar \Delta_T}{\bar C(\bar a)+\bar C_M}\,.}

\noindent The introduction of a loading device has a twofold effect: on one hand it better represents the real testing conditions, on the other hand it provides a test control that is intermediate between the unstable $P$-driven test and the stable $\Delta$-driven test. This results in a non-convex free energy function leading to an equilibrium states domain that includes an unstable branch for short cracks and a stable branch for longer cracks.

The data sets used in this paper are generated artificially to mimic different standard resistance models.

In the following, the data-driven results obtained using either Algorithm~\ref{algo:DD_GLOB} or Algorithms~\ref{algo:DD_DIST}-\ref{algo:DD_CONS} are compared with the reference solution of the global and local minimization problem, respectively. Global minimization is performed numerically at each time step by evaluating (\ref{eq:mixed_varpr-1})  at 1000 equispaced points along the interval $[0, \bar L=1]$ and finding the solution with (\ref{eq:glob-1}). The local minimizer is obtained analytically solving (\ref{eq:KT}). In this case the possible occurrence of crack jumps and their position is not uniquely defined. However, to determine them one can invoke causality or Onsager's principle. This discussion is out of the scope of this paper but the interested reader can refer to \cite{Negri2010,Alessi2016}. It is also worth mentioning that, for the case at hand, the globally stable states domain provides, in absence of initial defects, a finite crack nucleation load. Conversely, the local minimality condition forbids crack nucleation at finite loads in agreement with Griffith's theory.

\FloatBarrier
	\subsection{Griffith fracture} \label{sct:DD_Griff}

	In this section we present the results obtained using Griffith model, for which
	\eqn{eq:Griff_err}{\bar G_R = 1\,,}
\noindent and
	\eqn{eq:Griff_Fr}{\frac{\bar F_R}{\bar b} = \bar G_R \bar a = \bar a\,,}
\noindent in terms of critical energy release rate and critical energy, respectively.  The data sets used for the data-driven solution are obtained through a sampling of  (\ref{eq:Griff_err}) and (\ref{eq:Griff_Fr}) with 50 points randomly distributed along the interval $\bar a=[0, 1.1\bar L]$. To allow a fair comparison between the different approaches, (\ref{eq:Griff_err}) and (\ref{eq:Griff_Fr}) are sampled at the same values of $\bar a$.

		\subsubsection{Global minimization}  \label{sct:Griff_noNoise_glob}
\begin{figure}[!h]
	\begin{adjustwidth}{-3cm}{-3cm}
	\centering
		\includegraphics{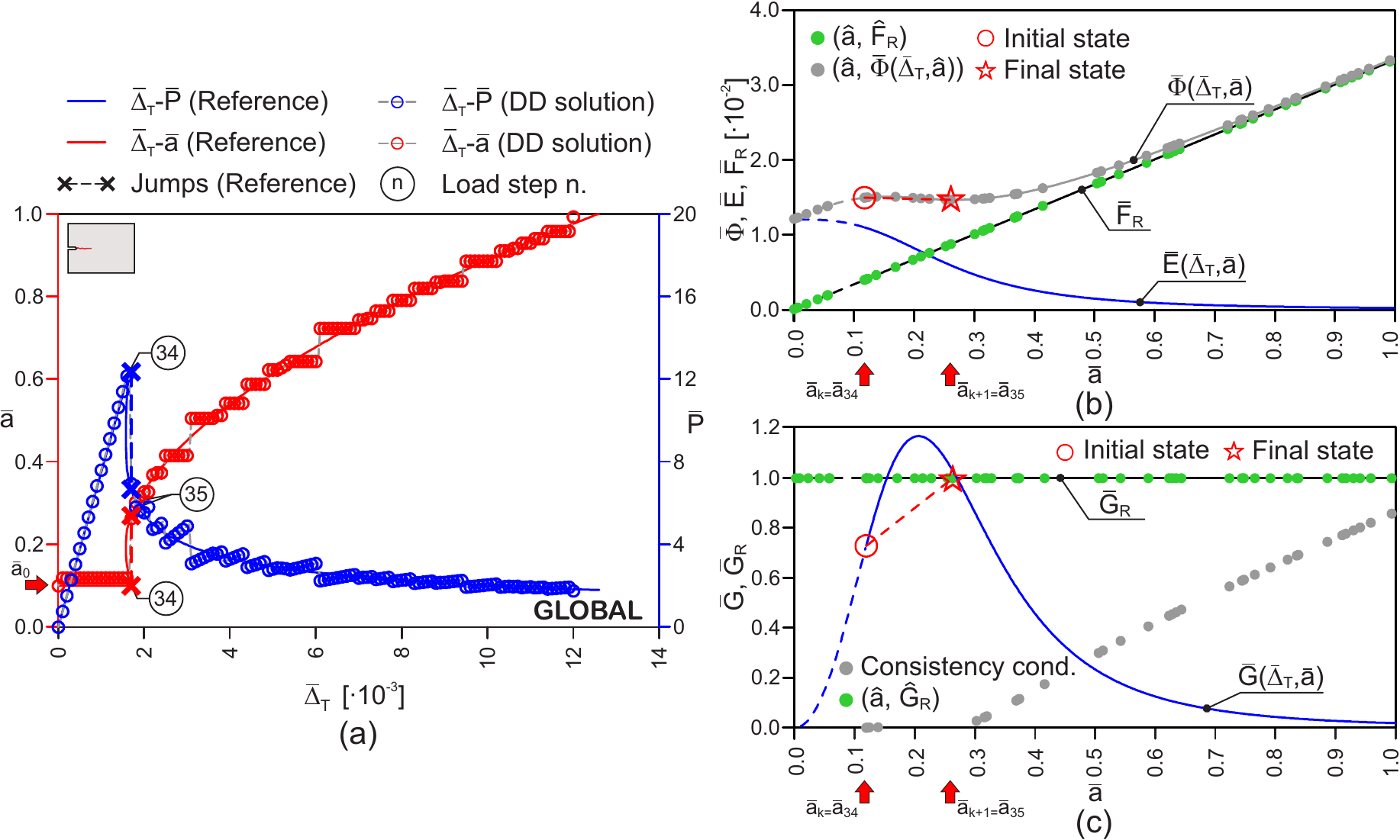}
	\end{adjustwidth}
		\caption{Data-driven results for Griffith fracture using global minimization: comparison between reference and data-driven results (a),  data-driven search procedure at load step 35 (i.~e.,at the crack jump) in terms of energy (b) and of energy release rate (c).}
		\label{fig:glob_sol}
	\end{figure}

Fig.~\ref{fig:glob_sol}a shows the comparison between the reference and data-driven (DD in the figures) solution obtained using the global minimization approach (Algorithm~\ref{algo:DD_GLOB}) and the material data set shown in Fig.~\ref{fig:glob_sol}b.

Considering the limited number of data points used and the presence of large unsampled intervals of crack length (Figs.~\ref{fig:glob_sol}b,c), the agreement between reference and data-driven solution is excellent although the crack length undergoes a small increment already at the first load step (Fig.~\ref{fig:glob_sol}a) which is not present in the reference solution. This happens because there is no material data point for a crack length equal to $\bar a_0$ and, thus, the first data-driven search returns the closest globally stable point fulfilling the irreversibility condition.

	Notably, the reference curve predicts  a crack jump at the load step 35, i.~e., for a machine displacement $\bar \Delta_T$ = 1.75$\cdot$10$^{-3}$, which is correctly reproduced by the data-driven solution in terms of both position and extension although no ad-hoc criterion is introduced in Algorithm~\ref{algo:DD_GLOB} (Fig.~\ref{fig:glob_sol}a). Fig.~\ref{fig:glob_sol}b shows the main energetic quantities at this load step, which can be taken as a paradigmatic example for the data-driven global minimization procedure. Here, it can be appreciated how the energy $\bar E(\bar \Delta_T, \bar a)$ and the resistance energy term $\bar F_R$ are composed in order to obtain the data set $\{\hat a_i, \bar\Phi(\bar\Delta_{T\,35},\hat a_i)\}$, on which minimization is performed to obtain the globally stable state at this load step. Figs. ~\ref{fig:glob_sol}a and ~\ref{fig:glob_sol}b show $\bar a_k=\bar a_{34}$, i.~e., the crack length at the previous load step 34, which corresponds to the initial unstable state, as well as the computed crack length solution for the current load step $\bar a_{k+1}=\bar a_{35}$.

 The crack jump in Fig.~\ref{fig:glob_sol}a is due to an unstable branch in the crack evolution curve which is the result of the non-convexity of the free-energy function. In particular, the initial value of the crack length $\bar a_0$ lies within an ascending branch of the free energy (Fig.~\ref{fig:glob_sol}a) part of which is inaccessible due to the irreversibility constraint (dashed line in Fig.~\ref{fig:glob_sol}b,c). This initial value corresponds to the global minimum of the accessible part of the function $\bar \Phi(\bar \Delta_T, \bar a)$ until the device displacement $\bar \Delta_T$ = 1.75$\cdot$10$^{-3}$, when the global minimum of the function suddenly changes its position from $\bar a = \bar a_0$ to $\bar a \simeq 0.27$. This forces the crack to extend until $\bar a \simeq 0.27$, which is seen by the system as a more favorable state. The finite increment of crack length induces an abrupt change in the compliance of the specimen triggering a snap back in the load-displacement curve (Fig.~\ref{fig:glob_sol}a). From a physical standpoint the crack jumps are the manifestation of dynamic effects not captured by the rate-independent quasi-static framework \cite{Negri2010b}. Fig.~\ref{fig:glob_sol}b shows also that, adopting a global minimality principle, in both data-driven and reference solution the state of the system at the crack jumps is allowed to overcome arbitrarily large energetic barriers violating the causality principle, as mentioned in sect.~\ref{subsec:cracked_solid}  and discussed in \cite{Negri2010,Alessi2016}.

For comparison purposes, the load step 35 is represented in terms of energy release rate in Fig.~\ref{fig:glob_sol}c, where we can observe that the initial state cannot be considered unstable from the standpoint of (\ref{eq:KT}). This is a direct consequence of the globally stable domain being smaller compared to the domain of the locally stable states, as discussed in \cite{Negri2010,Alessi2016}. Fig.~\ref{fig:glob_sol}c also reports the set of candidate points for the local minimum of Algorithm~\ref{algo:DD_CONS}  (set $\mathcal{D}_{R\,k+1}=\mathcal{D}_{R\,35}$ in Algorithm~\ref{algo:DD_CONS}). The solution point obtained with the global minimum algorithm is not included, i.~e., this point is not a candidate for Algorithm~\ref{algo:DD_CONS}, while it is a good candidate for Algorithm~\ref{algo:DD_DIST} as it is quite close to the curve $\bar G(\bar \Delta_T,\,\bar a)$.

\FloatBarrier

	\subsubsection{Local minimization} \label{sct:griff_locmin}
\begin{figure}[!h]
	\begin{adjustwidth}{-3cm}{-3cm}
	\centering
		\includegraphics{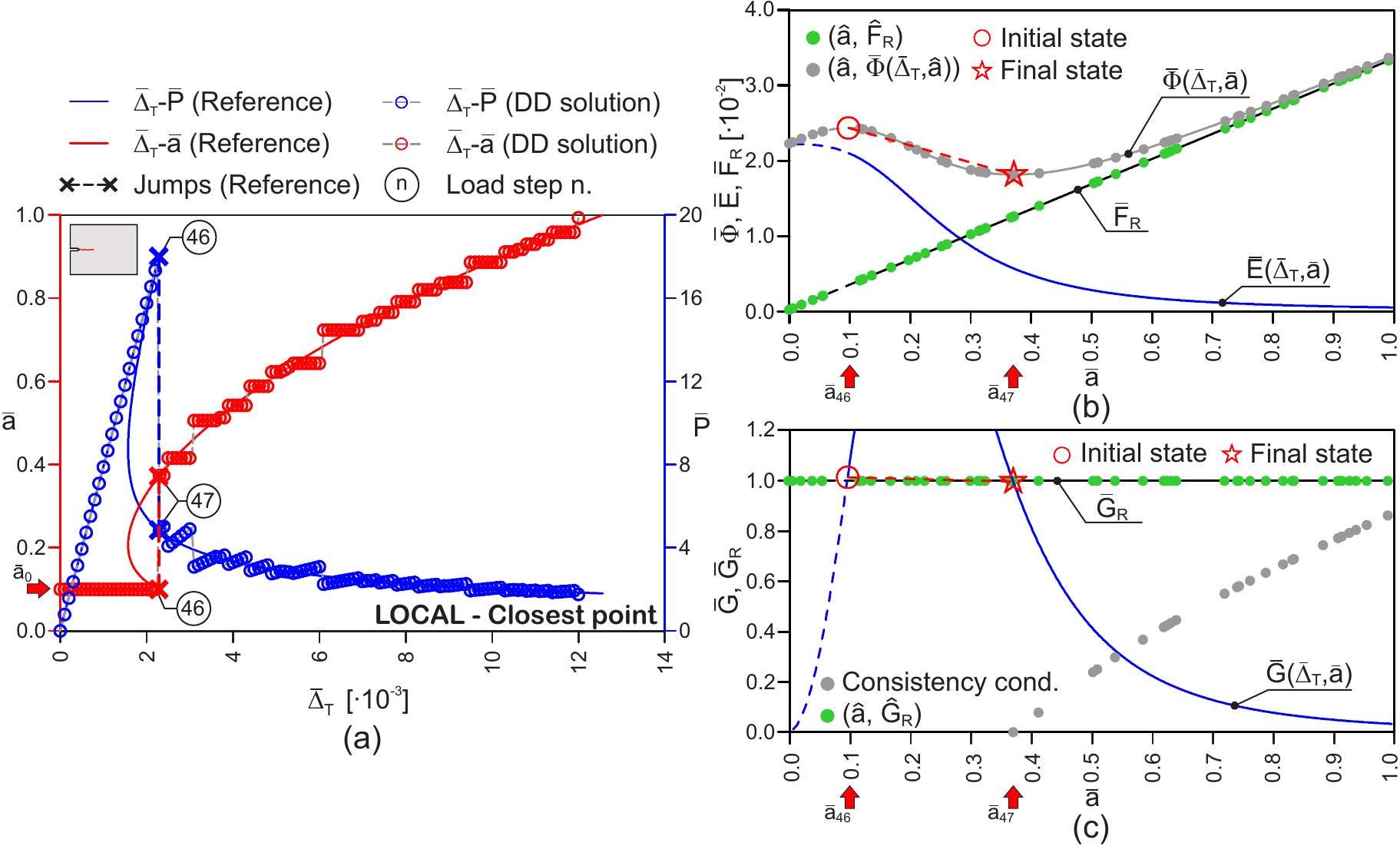}
	\end{adjustwidth}
		\caption{Data-driven results for Griffith fracture using local minimization and closest point projection: comparison between reference and data-driven results (a), data-driven search procedure at load step 47 (i.~e.,at the crack jump) in terms of energy (b) and of energy release rate (c).}
		\label{fig:dist_sol}
	\end{figure}

\begin{figure}[!h]
	\begin{adjustwidth}{-3cm}{-3cm}
	\centering
		\includegraphics{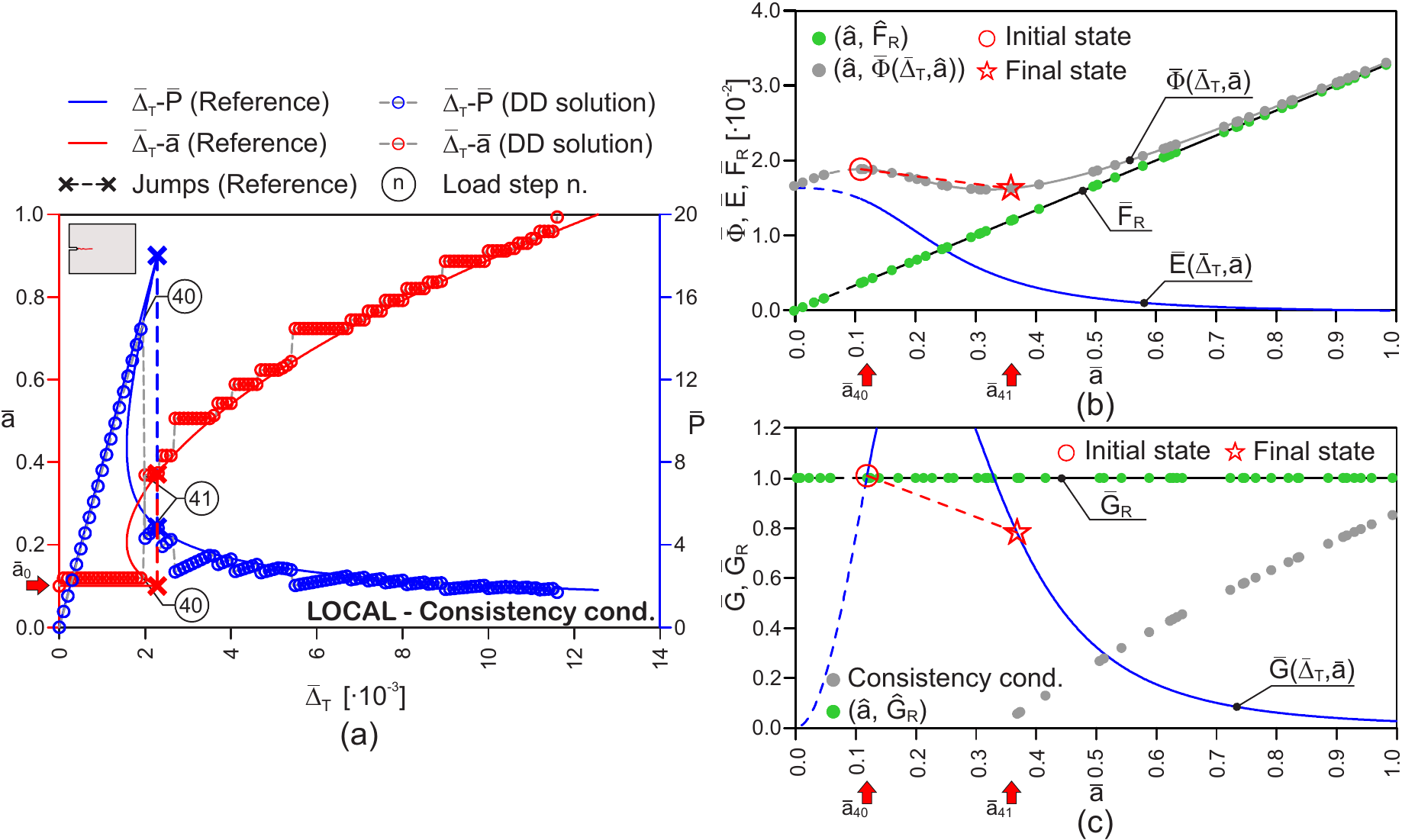}
	\end{adjustwidth}
		\caption{Data-driven results for Griffith fracture using local minimization and the consistency condition: comparison between reference and data-driven results (a), data-driven search procedure at load step 41 (i.~e.,at the crack jump) in terms of energy (b) and of energy release rate (c).}
		\label{fig:loc_sol}
	\end{figure}

	Figs.~\ref{fig:dist_sol}a and \ref{fig:loc_sol}a compare the reference and data-driven solutions obtained using local minimization along with closest point projection (\ref{data_loc-2}) and the consistency condition (\ref{data_loc-1-1}), respectively. In both cases the material data set is the same as in Sect.~\ref{sct:Griff_noNoise_glob}.

Also in this case the reference solution involves a crack jump that takes place at $\bar\Delta_T$ = 2.23$\cdot$10$^{-3}$. This value is higher than that obtained from global minimization, which is again consistent with the larger extension of the locally stable states domain \cite{Negri2008,Negri2010,Alessi2016}. Here the system does not evolve as long as there is an energetic barrier separating the current state and a more energetically favorable and accessible state. For a device displacement $\bar\Delta_T$ = 2.23$\cdot$10$^{-3}$ the state corresponding to $\bar a = \bar a_0$ is surrounded by accessible states characterized by lower energies. This makes the current state unstable and leads to the next quasi-static equilibrium condition at $\bar a  \simeq 0.37$, upon a finite crack size increment. This, once again, triggers a snap-back branch in the load-displacement curve. Also in this case dynamic effects can be advocated as underlying physical phenomenon that leads the system through a series of unstable states to a new (local) equilibrium condition \cite{Negri2010b}. The higher displacement attained before the crack jump also leads to a higher peak load, which is evident comparing Figs.~\ref{fig:dist_sol}a and \ref{fig:loc_sol}a with Fig.~\ref{fig:glob_sol}a.

Once again the agreement between reference and data-driven solution is excellent. The data-driven search procedure based on closest point projection performs better than that  based on the consistency condition, especially in predicting the peak load (Figs.~\ref{fig:dist_sol}a and \ref{fig:loc_sol}a). This is mainly due to the fact that the latter undergoes a limited evolution of the crack length at the first load step (Figs.~\ref{fig:loc_sol}a) for the same reason highlighted for the global minimum solution in Sect.~\ref{sct:Griff_noNoise_glob}. This does not hold for the closest point projection procedure (Figs.~\ref{fig:dist_sol}a) because of the initialization of $\bar G_R$ (Sect.~\ref{sct:DD_CPP}), which is mostly responsible for the first evolution of the crack length. For this reason the two approaches predict the crack jump to take place at two different load steps, namely at the 47$^{th}$ ($\bar\Delta_T$  = 2.35$\cdot$10$^{-3}$) and 41$^{st}$ ($\bar\Delta_T$ = 2.05$\cdot$10$^{-3}$) for the closest point projection and the consistency condition, respectively.

Figs.~\ref{fig:dist_sol}b,c and \ref{fig:loc_sol}b,c illustrate the data-driven search procedure at the crack jump for the two approaches. Apart from the peak load, results are very similar. In particular, they are both consistent with the characteristic of local minimality of preventing the system from overcoming energetic barriers at the crack jump (Figs.~\ref{fig:dist_sol}b and \ref{fig:loc_sol}b). The main difference is that the consistency condition strategy strictly enforces (\ref{eq:KT}b), hence, it has the tendency to overestimate the crack length and to accept values of the energy release rate smaller than the critical one at the final state (Fig.~\ref{fig:loc_sol}c). Conversely, the closest point projection allows also values of energy release rate slightly larger than the critical one, hence, it alternates between over- and underestimations of the crack length (Fig.~\ref{fig:dist_sol}c).

\FloatBarrier

	\subsubsection{Results in presence of noisy data}
	
\begin{figure}[!ht]
	\begin{adjustwidth}{-3cm}{-3cm}
	\centering
		\includegraphics{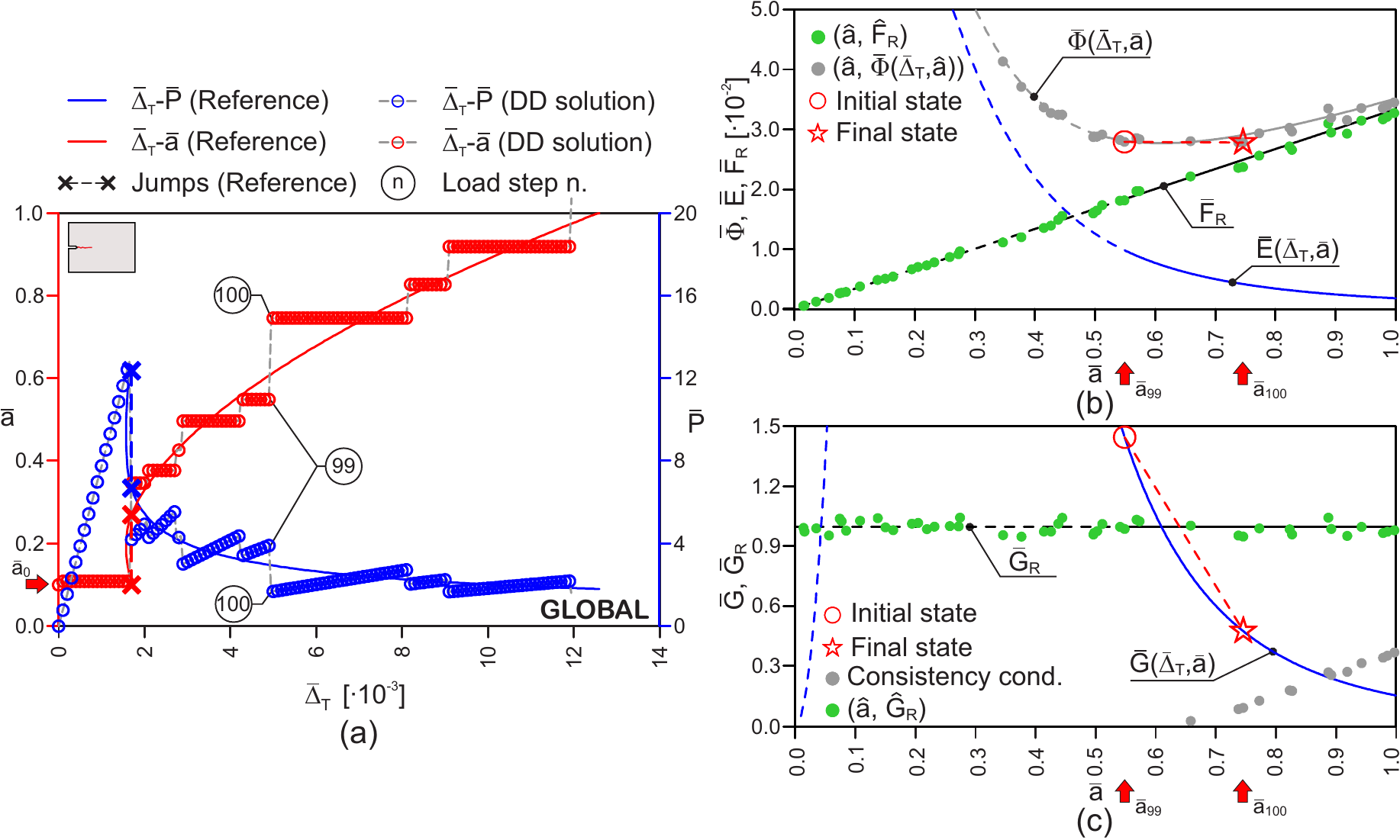}
	\end{adjustwidth}
		\caption{Data-driven results for Griffith fracture with a noisy material data set using global minimization: comparison between reference and data-driven results (a),  data-driven search procedure at load step 100 in terms of energy (b) and of energy release rate (c).}
		\label{fig:glob_sol_noise}
	\end{figure}

\begin{figure}[!ht]
	\begin{adjustwidth}{-3cm}{-3cm}
	\centering
		\includegraphics{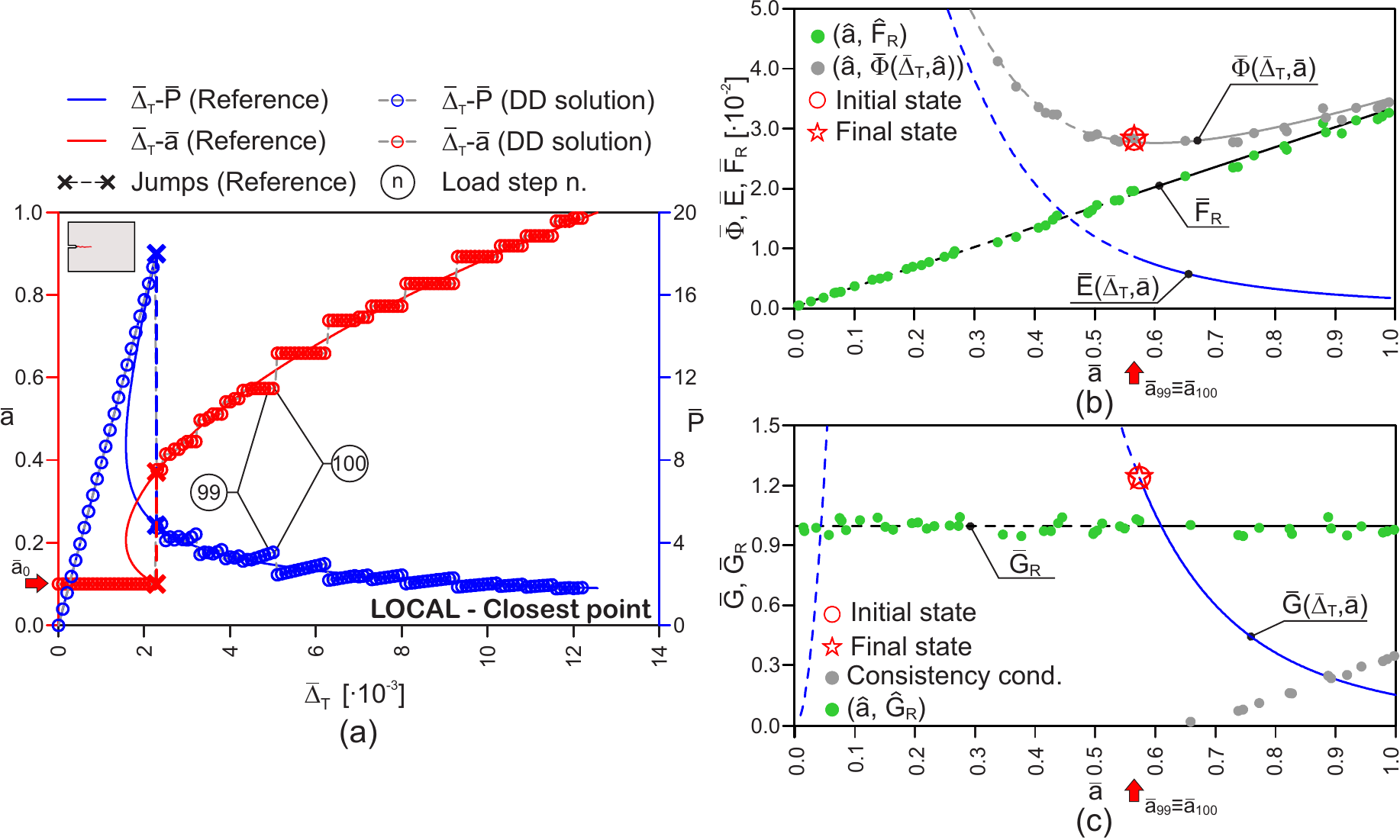}
	\end{adjustwidth}
		\caption{Data-driven results for Griffith fracture with a noisy material data set using local minimization and closest point projection: comparison between reference and data-driven results (a), data-driven search procedure at  load step 100 in terms of energy (b) and of energy release rate (c).}
		\label{fig:dist_sol_noise}
	\end{figure}

\begin{figure}[!hb]
	\begin{adjustwidth}{-3cm}{-3cm}
	\centering
		\includegraphics{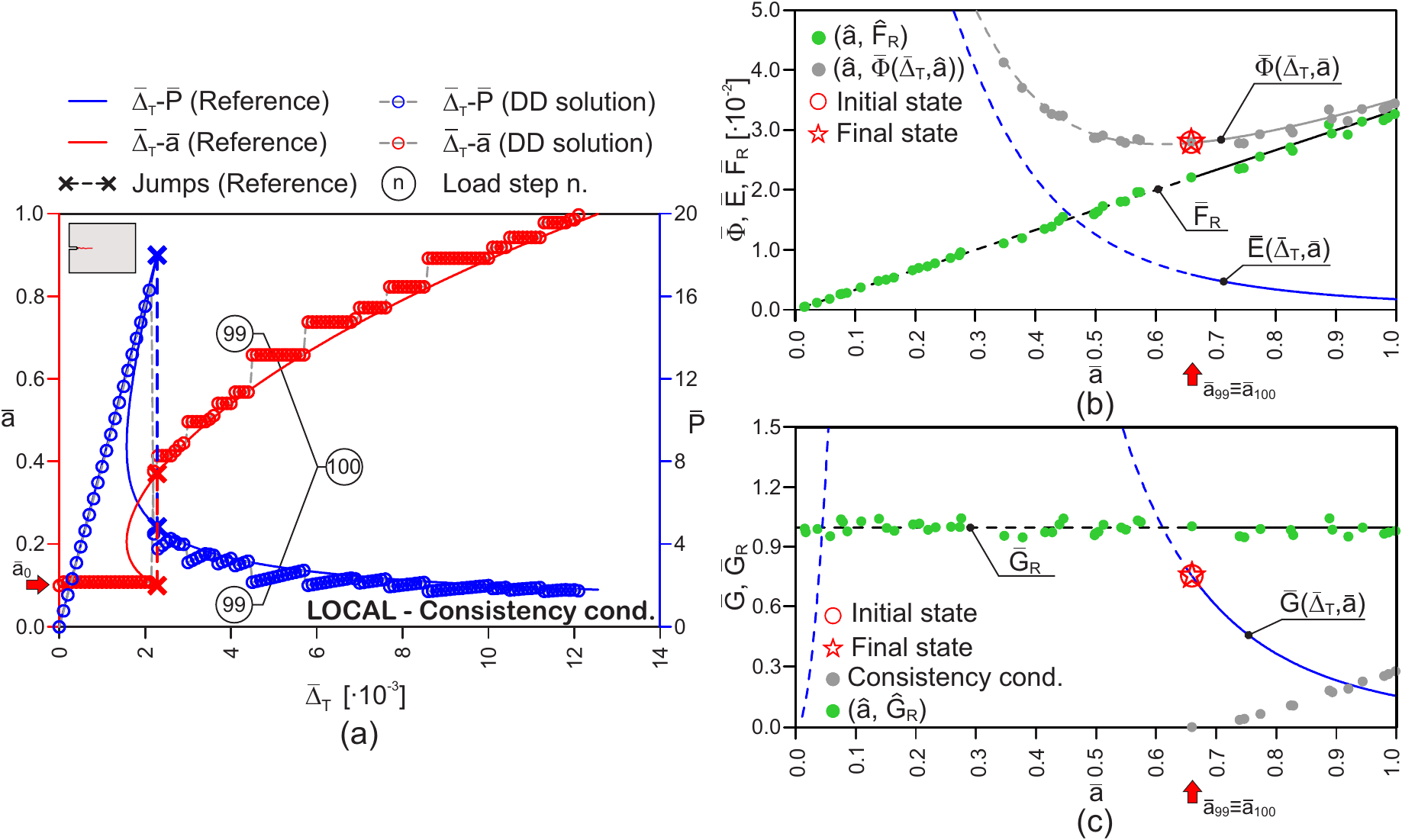}
	\end{adjustwidth}
		\caption{Data-driven results for Griffith fracture with a noisy material data set using local minimization and the consistency condition: comparison between reference and data-driven results (a), data-driven search procedure at load step 100 in terms of energy (b) and of energy release rate (c).}
		\label{fig:loc_noise}
	\end{figure}

	To determine the sensitivity of the proposed method to noise in the input data, a random white noise with amplitude $\pm$ 2.5\% in terms of difference between the observed and expected value is applied to the material data sets obtained by sampling (\ref{eq:Griff_err}) and (\ref{eq:Griff_Fr}). As before, sampling is performed at the same 50 values of crack lengths. Moreover, for consistency of both noisy data sets, points corresponding to the same values of crack length are affected by the same noise, i.e., for the same $\hat a_i$, the difference between expected and observed value of energy and energy release rate is the same. The results are illustrated in Figs.~\ref{fig:glob_sol_noise} to \ref{fig:loc_noise}.

Comparing the results with and without noise, it is clear that data-driven global minimization is very sensitive to the presence of noise. The displacement vs. crack length curve involves several large crack jumps followed by long elastic branches (Fig.~\ref{fig:glob_sol_noise}a). The data-driven search procedure related to the jump at load step 100 ($\bar \Delta_T$ = 5$\cdot$10$^{-3}$) with global minimization is illustrated in Fig.~\ref{fig:glob_sol_noise}b. Here the solution is seen to switch between points affected by large values of negative noise. In this case, this leads to an overestimation of the crack length delivering a solution point that refers clearly to an ascending branch of the reference noiseless total energy curve. From the standpoint of the energy release rate (Fig.~\ref{fig:glob_sol_noise}c), this means that the solution often evolves from points clearly above to points clearly below the critical energy release rate.	

The agreement between reference and data-driven results obtained with both local minimization procedures (Figs.~\ref{fig:dist_sol_noise}a and \ref{fig:loc_noise}a) is still excellent and comparable to that obtained in the noiseless case (Figs.~\ref{fig:dist_sol}a and \ref{fig:loc_sol}a). The same observations of Sect.~\ref{sct:griff_locmin} apply. Figs.~\ref{fig:dist_sol_noise}b,c and \ref{fig:loc_noise}b,c illustrate the data-driven search procedures and the corresponding energetic quantities at the same load step analyzed for global minimization (Fig.~\ref{fig:glob_sol_noise}b,c). Here we can see that the crack length for both local minimization strategies does not evolve. This is due to a relatively large unsampled interval of crack length, leading to a lack of better solution candidates than the values at load step 99. While with closest point projection the data-driven solution underestimates the reference solution (Fig.~\ref{fig:dist_sol_noise}a) because the energy release rate curve lies between two points of the data set (Fig.~\ref{fig:dist_sol_noise}c), the consistency condition criterion overestimates the crack length (Fig.~\ref{fig:loc_noise}a,c). Nevertheless, in both cases the solution is close to the (local) minimum of the total energy (Figs.~\ref{fig:dist_sol_noise}b and \ref{fig:loc_noise}b).

\FloatBarrier

\subsection{R-curve fracture}

This section shows that the algorithms presented in Sect.~\ref{sct:NUM_algo} can be applied without any modification to other fracture models. We adopt the following R-curve model

\eqn{eq:Rcurve}{\bar G_R(\bar a)=1+\frac{(\bar a -0.1)^2}{(\bar a -0.1)^2+0.2(\bar a -0.1)}\,,}

\noindent or, in terms of energy

\eqn{eq:Rcurve_FR}{\frac{\bar F_R(\bar a)}{\bar b} = 0.2\left[10\bar a - \text{ln}\,\left(\frac{\bar a +0.1}{0.1}\right)\right]\,.}

\begin{figure}[!hb]
	\begin{adjustwidth}{-3cm}{-3cm}
	\centering
		\includegraphics{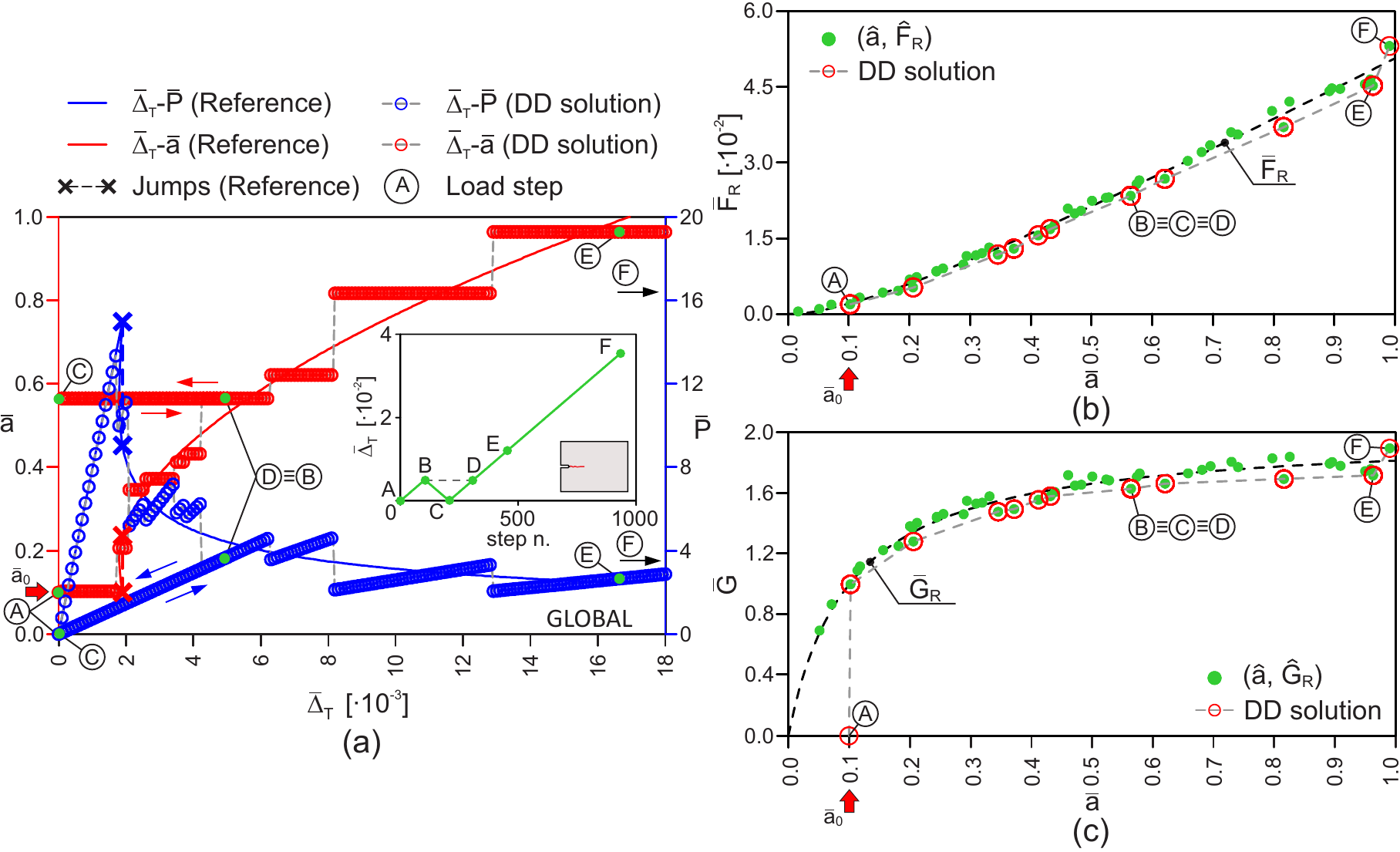}
	\end{adjustwidth}
		\caption{Data-driven results for an R-curve model with a noisy material data set using global minimization: comparison between reference and data-driven results (a), data-driven solution in terms of energy (b) and of energy release rate (c).}
		\label{fig:glob_rcurve}
	\end{figure}

\begin{figure}[!hb]
	\begin{adjustwidth}{-3cm}{-3cm}
	\centering
		\includegraphics{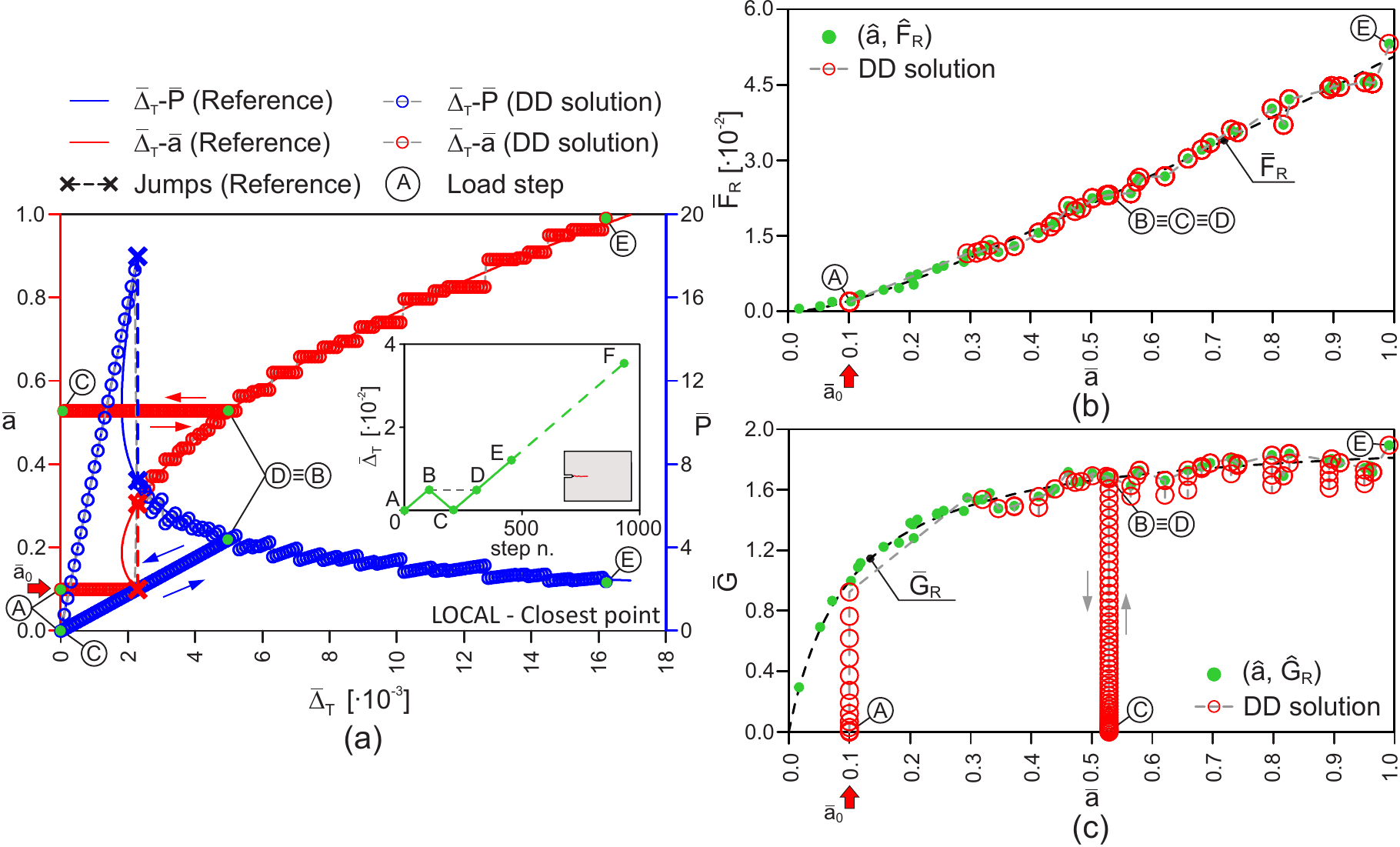}
	\end{adjustwidth}
		\caption{Data-driven results for an R-curve model with a noisy material data set using local minimization and closest point projection: comparison between reference and data-driven results (a), data-driven solution in terms of energy (b) and of energy release rate (c).}
		\label{fig:dist_rcurve}
	\end{figure}

Unless otherwise specified here and in the following, a random noise with amplitude $\pm$2.5\% is added to the material data set, which is composed of 50 randomly distributed points. Moreover, the loading ramp now includes a complete unloading-reloading branch that starts at $\bar \Delta_T$ = 5$\cdot$10$^{-3}$. The results obtained with global minimization are illustrated in Fig.~\ref{fig:glob_rcurve}, while those of local minimization with closest point projection are presented in Fig.~\ref{fig:dist_rcurve}. From now on, the local minimization results obtained with the consistency condition are not reported, as they are similar to those obtained with closest point projection.

Both approaches are able to correctly follow the unloading-reloading curve $\textsf{A-B-C-D}$, see Figs.~\ref{fig:glob_rcurve}a and \ref{fig:dist_rcurve}a. In particular, the unloading  and reloading phases $\textsf{B-C}$ and $\textsf{C-D}$ follow a linear elastic path along the displacement vs. load curve, while the crack length remains constant (Figs.~\ref{fig:glob_rcurve}b,c and \ref{fig:dist_rcurve}b,c).

The agreement between reference and data-driven results is similar as in the example with Griffith fracture. In particular, the presence of noise significantly affects the accuracy of the global minimization approach (Fig.~\ref{fig:glob_rcurve}a), as evident e.g. in the post peak portion of $\textsf{A-B}$ in Fig.~\ref{fig:glob_rcurve}a. Figs.~\ref{fig:glob_rcurve}b,c show that among all the available states, global minimization tends to prefer points affected by a negative noise, thus promoting lower total energy (Figs.~\ref{fig:glob_rcurve}b). In the current example this causes the data-driven procedure to highly overestimate the ultimate displacement, defined as the displacement corresponding to $\bar a$ = 1. The reference solution predicts an ultimate displacement $\bar \Delta_T$ = 16.9$\cdot$10$^{-3}$ while global minimization reaches the ultimate condition at $\textsf{F}$ for $\bar \Delta_T$ = 35.3$\cdot$10$^{-3}$ (the last portion of the curves is not shown in Fig.~\ref{fig:glob_rcurve}a). This is due to the negative noise that affects the data-driven solution point $\textsf{E}$ preceding the one leading to the ultimate condition (Figs.~\ref{fig:glob_rcurve}b,c). The ultimate point $\textsf{F}$ is affected by a positive noise. Hence, to allow the ultimate state to become a global minimum, a large amount of energy must be provided to the system. Conversely, the local minimization approach is not biased towards any specific set of points and can switch to the ultimate condition much earlier, namely already at point $\textsf{E}$ (Figs.~\ref{fig:dist_rcurve}b,c).

As mentioned earlier, local minimization based on closest point projection is the only data-driven procedure able to differentiate between dissipative and elastic steps. Such distinction is visible in Fig.~\ref{fig:dist_rcurve}c, where the initial and unloading-reloading linear elastic phases are evident.

\FloatBarrier
\newpage

\subsection{Bimaterial DCB}

\begin{figure}[!h]
	\begin{adjustwidth}{-3cm}{-3cm}
	\centering
		\includegraphics{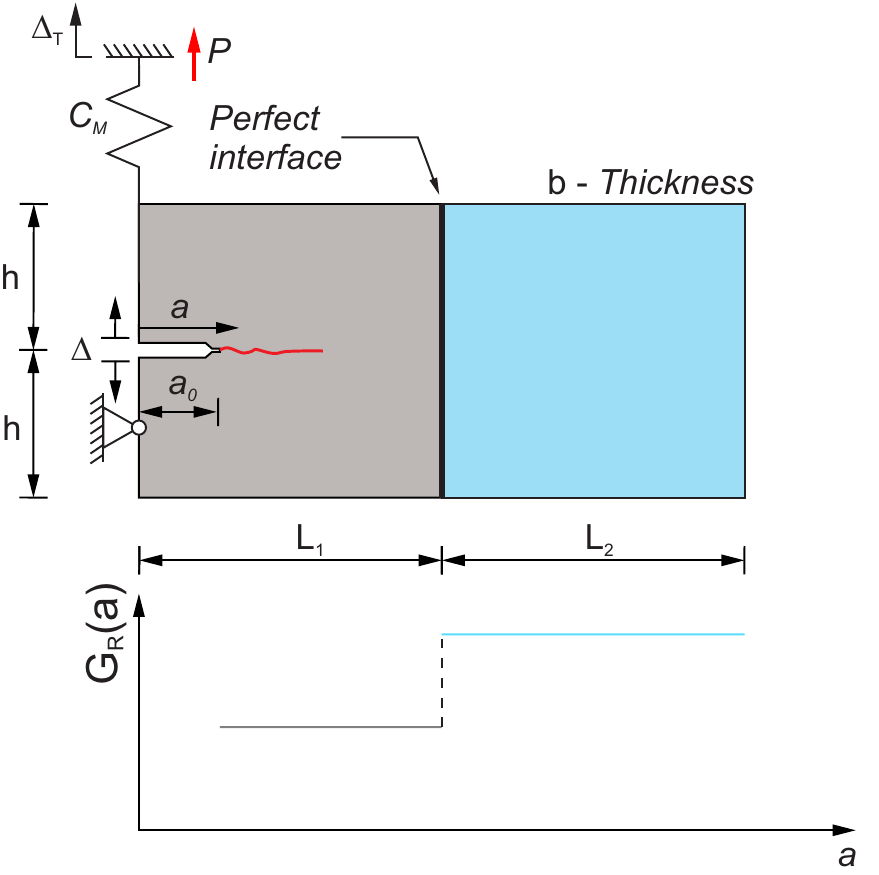}
	\end{adjustwidth}
		\caption{Sketch of the bimaterial DCB specimen.}
		\label{fig:interf}
	\end{figure}

Let us now consider a DCB specimen composed of two different materials connected by a perfect interface (Fig.~\ref{fig:interf}). The lengths of the two parts of the sample are $\bar L_1=\bar L_2 = 0.5$ ($\bar L_i=L_i/L$). We model Griffith fracture with

\eqn{eq:err_intstab}{\bar G_R(\bar a)=\begin{cases} 1 &\text{for } 0.0< \bar a \le 0.5\\
														 5 & \text{for } 0.5< \bar a \le 1.0\,,  \end{cases}}

\noindent or, in terms of energy

\eqn{eq:FR_intstab}{\frac{\bar F_R(\bar a)}{\bar b} = \begin{cases} \bar a &\text{for } 0.0< \bar a \le 0.5\\
														 (5\bar a-2) & \text{for } 0.5< \bar a \le 1.0\,.  \end{cases}}

\noindent In this case, the crack must first traverse the weaker half of the sample to reach the strongest part, therefore this arrangement is termed \emph{stable}. One can also define an \emph{unstable} setup where the materials are reversed, i.~e., \eqn{eq:err_intunstab}{\bar G_R(\bar a)=\begin{cases} 5 &\text{for } 0.0< \bar a \le 0.5\\
														 1 & \text{for } 0.5< \bar a \le 1.0\,,  \end{cases}}

\noindent or, in terms of energy

\eqn{eq:FR_intunstab}{\frac{\bar F_R(\bar a)}{\bar b} = \begin{cases}5 \bar a &\text{for } 0.0< \bar a \le 0.5\\
														 (\bar a+2) & \text{for } 0.5< \bar a \le 1.0\,.  \end{cases}}

\begin{figure}[!hb]
	\begin{adjustwidth}{-3cm}{-3cm}
	\centering
		\includegraphics{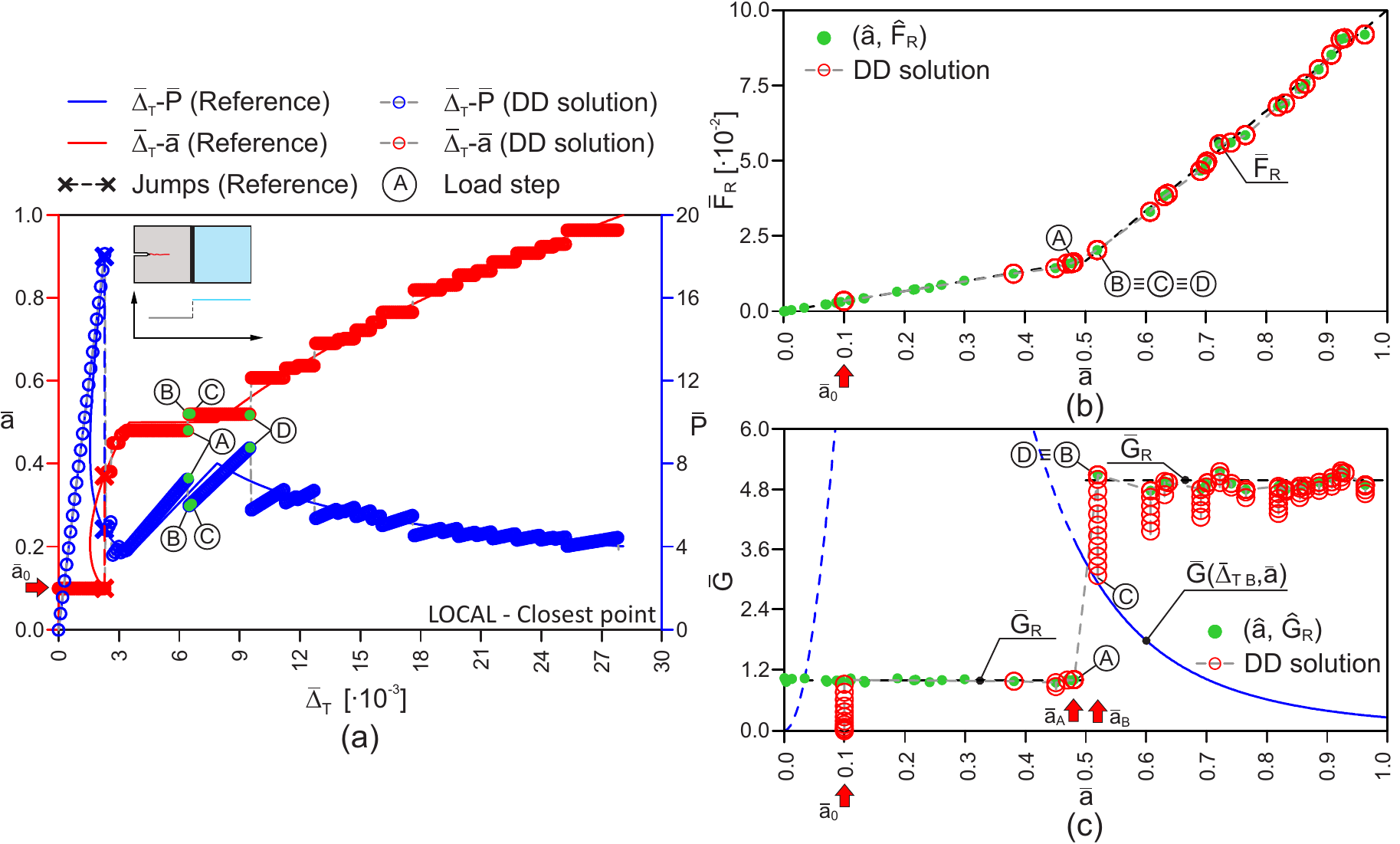}
	\end{adjustwidth}
		\caption{Data-driven results for the bimaterial DCB specimen (stable arrangement) using local minimization and closest point projection: comparison between reference and data-driven results (a), data-driven solution in terms of energy (b) and of energy release rate (c).}
		\label{fig:dist_interf_stab}
	\end{figure}

Because of the drawbacks of global minimization discussed in the previous examples, from now on only local minimization is pursued. In Fig.~\ref{fig:dist_interf_stab}a the reference and data-driven results obtained for the stable arrangement are compared and the agreement is very good. From the reference curves we can see that the crack stops propagating once it reaches the interface at $\bar a$ = 0.5, until the energy release rate reaches the critical value of the strongest material. The data-driven solution is able to reproduce this behavior, however, due to the discrete set of states in the material data, the transition between the two halves of the specimen occurs approximately half way on the horizontal branch of the reference crack length vs. displacement curve, i.~e., between the load steps $\textsf{A}$ and $\textsf{B}$ in Fig.~\ref{fig:dist_interf_stab}. At this stage of the test the energy release rate curve is approximately at the same distance from the last available material point of the weakest and the first one of the strongest material (Fig.~\ref{fig:dist_interf_stab}c). Since the transition occurs when the energy release rate is below the largest critical value, at the following load step $\textsf{C}$ a linear elastic branch starts that persists until the evolution conditions are met again at load step $\textsf{D}$ (Fig.~\ref{fig:dist_interf_stab}b,c).

\begin{figure}[!hb]
	\begin{adjustwidth}{-3cm}{-3cm}
	\centering
		\includegraphics{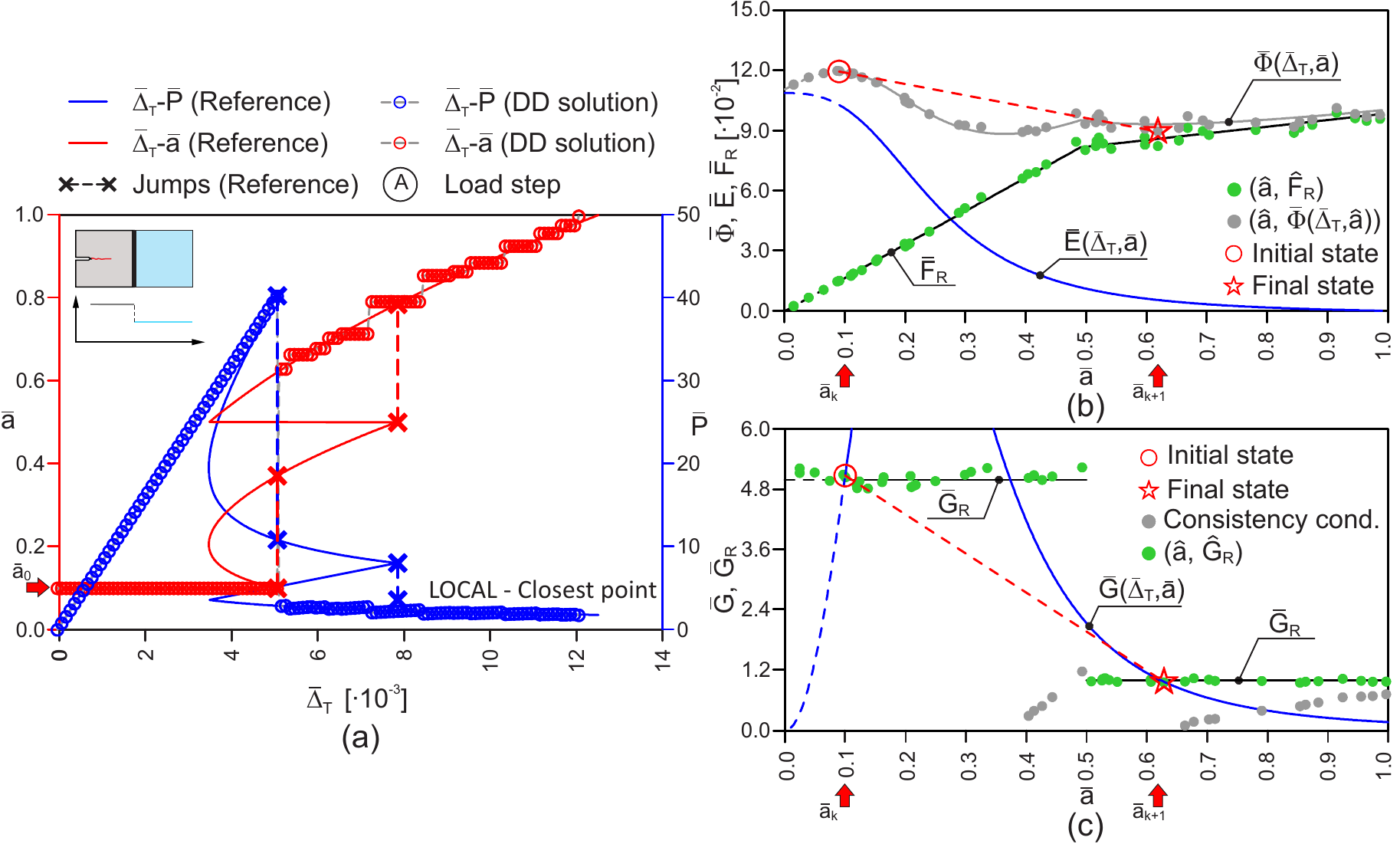}
	\end{adjustwidth}
		\caption{Data-driven results for the bimaterial DCB specimen (unstable arrangement) using local minimization and closest point projection: comparison between reference and data-driven results (a), data-driven search procedure at the load step corresponding to the crack jump in terms of energy (b) and of energy release rate (c).}
		\label{fig:dist_interf_unstab}
	\end{figure}

We now consider the \emph{unstable} setup, whose results are presented in Fig.~\ref{fig:dist_interf_unstab}. The reference curves display two snap-backs, the first one at the peak load and the second one at $\bar a$ = 0.5, i.~e., when the crack tip reaches the interface. As shown in Fig.~\ref{fig:dist_interf_unstab}a the data-driven approach does not reproduce the two jumps corresponding to these snap-backs, but merges them into one large jump. After reaching the peak load the crack tip jumps directly to the weakest portion of the specimen, then it follows again the reference solution. This result is due to a competition between different possible states after the peak load, as illustrated in Fig.~\ref{fig:dist_interf_unstab}b,c. For the analytical solution one can invoke causality or Onsager's principle \cite{Negri2010} to favor the higher branch of the reference solution after the peak over the lower branch taken directly by the data-driven response. However, these considerations play no role in the data-driven procedures (as a side note, similar results are obtained with global minimization and with local minimization using a consistency condition criterion - not shown).

\FloatBarrier

\subsection{Tapered DCB}

\begin{figure}[!h]
	\begin{adjustwidth}{-3cm}{-3cm}
	\centering
		\includegraphics{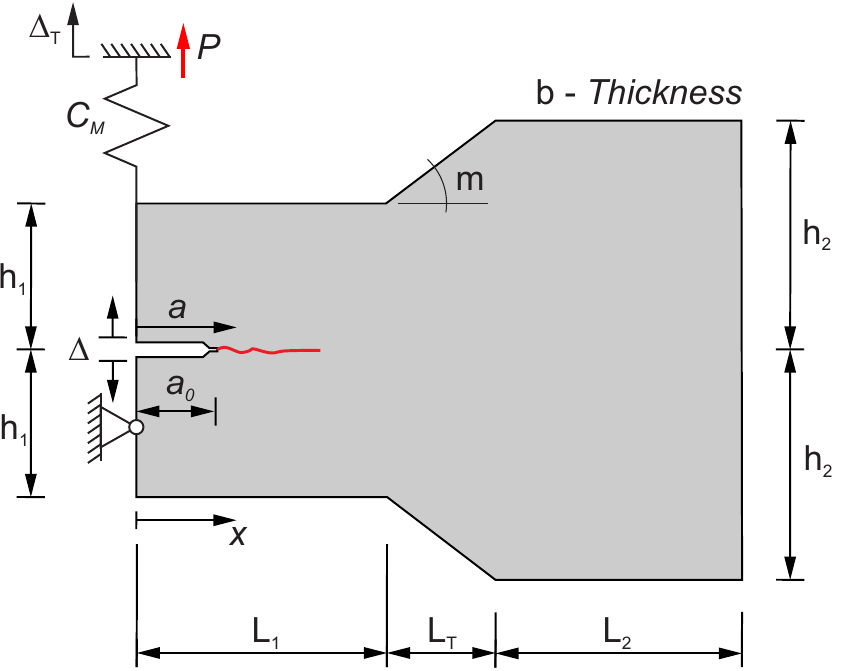}
	\end{adjustwidth}
		\caption{Sketch of the tapered DCB specimen.}
		\label{fig:tapered}
	\end{figure}

In this section we analyze a homogeneous tapered DCB specimen, see Fig.~\ref{fig:tapered}. We assume the same Griffith model of Sect.~\ref{sct:DD_Griff}, and the data set is composed of 100 points affected by random noise with amplitude $\pm 2.5\%$.

Unlike in the bimaterial DCB, here a transition in the cracking behavior is induced by the geometry of the specimen. The analytical compliance function is reported in Appendix~\ref{app:taper_comp}. As follows, three geometries are investigated, one with increasing height, and two with decreasing height and different slopes of the transition region, see Tab.~\ref{tab:taper_param}.
	
\begin{table}
\centering
\begin{tabular}{cccc}
\toprule
{\bf Parameter} & {\bf{\textsf{CASE 1}}} & {\bf{\textsf{CASE 2}}}& {\bf{\textsf{CASE 3}}}\\[4pt]
\toprule
 $\bar h_1$  & 0.10 & 0.10 & 0.10 \\[4pt]
 $\bar h_2$  & 0.15 & 0.05 & 0.04 \\[4pt]
 $\bar L_1$  & 0.50 & 0.45 & 0.45   \\[4pt]
 $\bar L_T$  & 0.10 & 0.30 & 0.10   \\[4pt]
 $\bar L_2$  & 0.40 & 0.25 & 0.45  \\[4pt]
 $m$ ($^\circ$)           & 1/2 (26.56$^\circ$) & -1/6 (-9.46$^\circ$) & -3/5 (-30.96$^\circ$) \\
\toprule
\end{tabular}
\caption{Parameters for the tapered DCB.}
\label{tab:taper_param}
\end{table}

\begin{figure}[!hb]
	\begin{adjustwidth}{-3cm}{-3cm}
	\centering
		\includegraphics{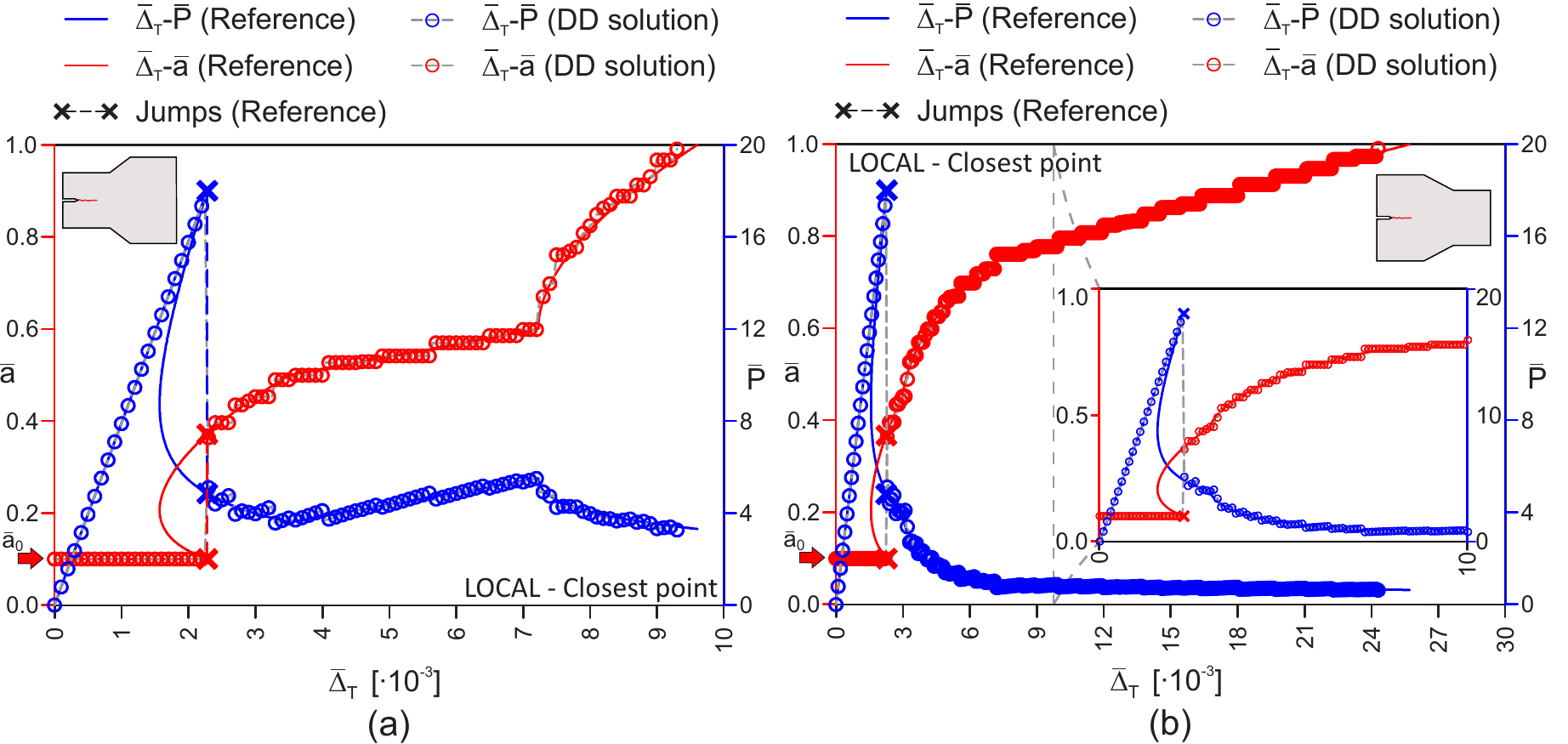}
	\end{adjustwidth}
		\caption{Comparison between reference and data-driven results for the tapered DCB specimen: $\textsf{CASE 1}$ (a) and $\textsf{CASE 2}$ (b).}
		\label{fig:taper_stab}
	\end{figure}

\begin{figure}[!hb]
	\begin{adjustwidth}{-3cm}{-3cm}
	\centering
		\includegraphics{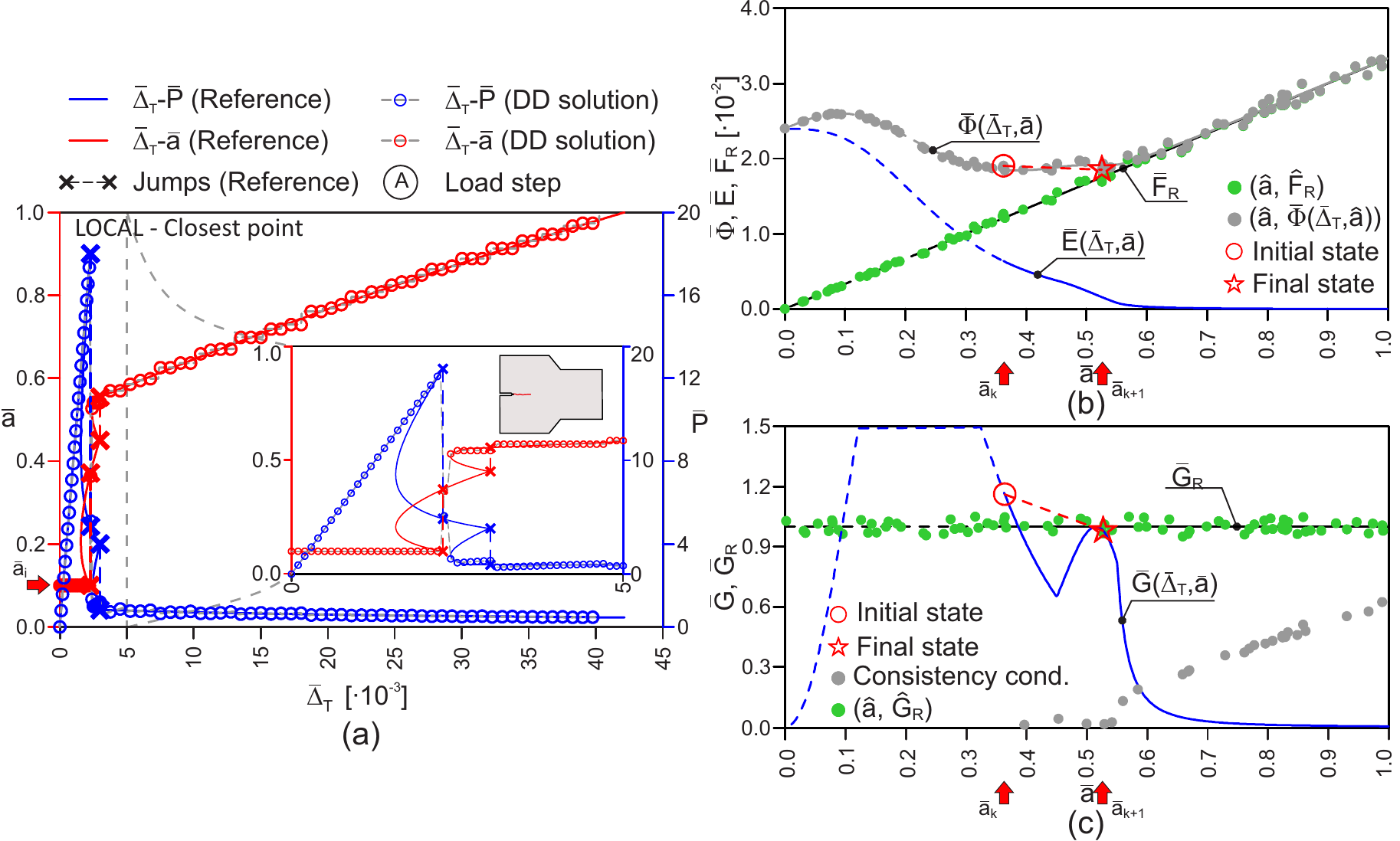}
	\end{adjustwidth}
		\caption{Data-driven results for the tapered DCB specimen ($\textsf{CASE 3}$) using local minimization and closest point projection: comparison between reference and data-driven results (a), data-driven search procedure for the load step corresponding to the second crack jump in terms of energy (b) and of energy release rate (c).}
		\label{fig:taper_unstab}
	\end{figure}

	As shown in Fig.~\ref{fig:taper_stab}, for the first two cases the results are similar as for the standard DCB. The reference solution features a snap-back whose position and extension is correctly reproduced by the data-driven results. In the third case the reference solution displays a second snap-back that takes place for $\bar a$ = $\bar L_1$. As for the bimaterial DCB with the unstable arrangement, the position of the second jump is clearly anticipated by the data-driven procedure (Fig.~\ref{fig:taper_unstab}a). This is caused once again by the competition of different possible solution states, see Figs.~\ref{fig:taper_unstab}b,c. After the second jump, the data-driven procedure goes back to closely reproducing the reference solution (Fig.~\ref{fig:taper_unstab}a).

	\subsection{Convergence }

Aim of this section is to demonstrate convergence of the data-driven results to the reference solution with respect to the number of points in the material data set and the noise amplitude for both global and local minimization algorithms. To this end, we consider a DCB test for a Griffith material as in Sect.~\ref{sct:DD_Griff} and we compute the error $\varepsilon$ of the data-driven solution with respect to the reference solution as

\eqn{eq:error}{\varepsilon = \sqrt{\sum_{k=1}^N\left(\bar a_k-\bar a^{ref}_k\right)^2}\,,}

\noindent where $\bar a_k$ and $\bar a^{ref}_k$ are respectively the data-driven and the reference solution computed for the same value of machine displacement $\bar \Delta_{T\,k}$. For each case considered in the following, we compute the average $\mu$ and the standard deviation of the error for 100 solutions obtained changing the randomly sampled material database and, if applicable, the random noise. We also compute the frequency histograms of the error (in the range $[0,\, 2\mu]$ with bin size of 0.1$\mu$) and the data-driven solution range of the $\bar a$ vs. $\bar \Delta_T$ curve. The latter is the smallest area in the $\bar \Delta_T-\bar a$ plane that includes all the 100 data-driven solutions. The load step size is the same adopted in the previous sections, namely $\bar \delta_T$ = 5$\cdot$10$^{-5}$.

\begin{figure}[!hb]
	\begin{adjustwidth}{-3cm}{-3cm}
	\centering
		\includegraphics{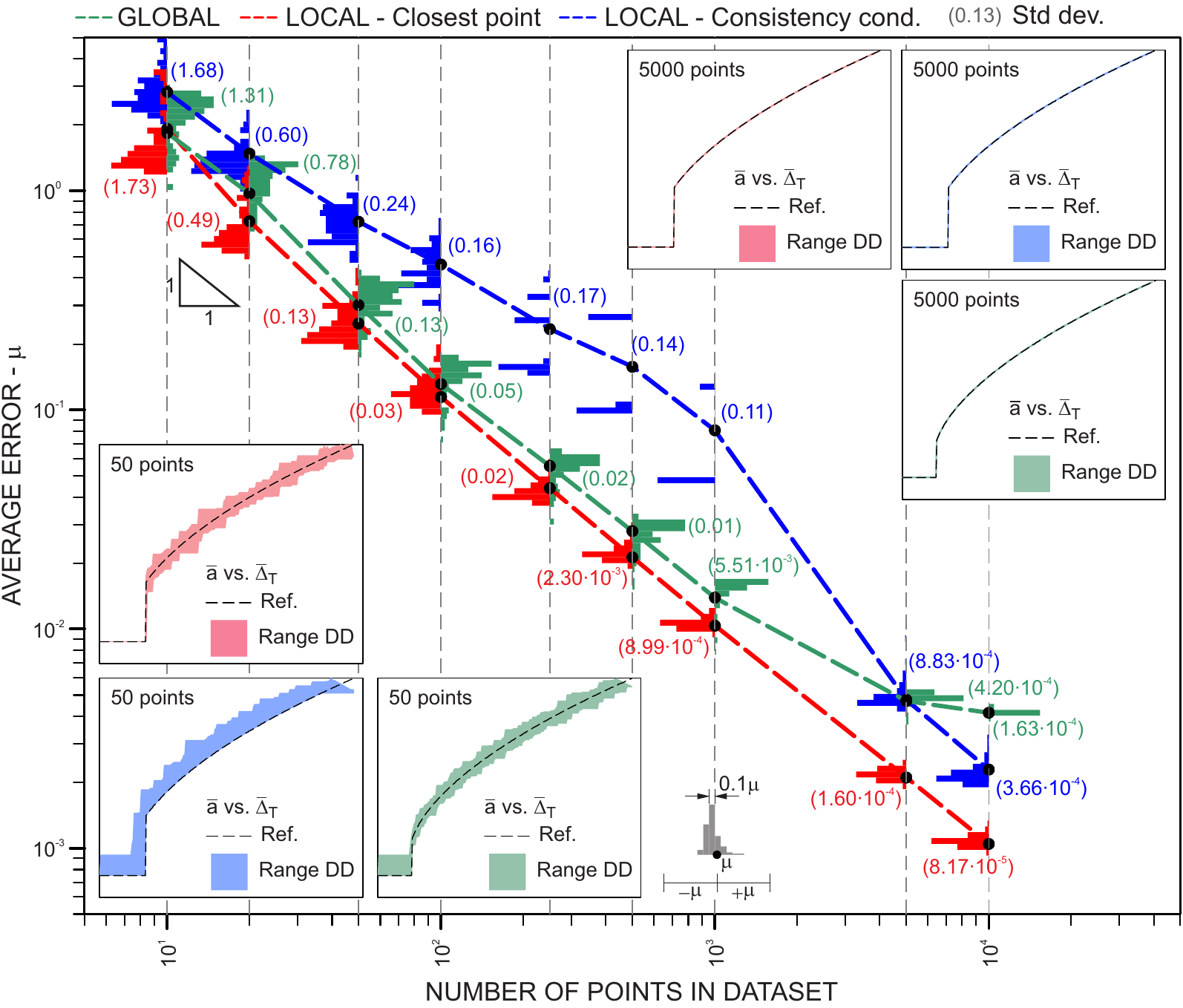}
	\end{adjustwidth}
		\caption{Convergence with respect to the number of points in the dataset including frequency histograms of the error and range of the solutions for 50 and 5000 points.}
		\label{fig:conv_pts}
	\end{figure}

Fig.~\ref{fig:conv_pts} show the results of the convergence study with respect to the number of points in a noiseless dataset. The average error is reported for 10, 20, 50, 100, 250, 500, 1000, 5000, and 10000 points. We can observe that for all approaches the rate of convergence is almost linear, with the closest point projection algorithm performing better than the other two. The consistency condition approach features an initially lower convergence rate, due to a relatively high error contribution given by the crack propagation at the first load step when the initial crack length is not (even approximately) resolved within the dataset. This causes a delayed or anticipated initial crack jump responsible for a major contribution to the error, whose amplitude is governed by the snap-back characteristics and roughly independent of the dataset properties. Because of this phenomenon, increasing the size of the dataset leads to the clustering of the solutions in two groups: one featuring an incorrectly predicted crack jump affected by a higher error and one where the initial crack length is closely resolved and the error is much lower. This can be clearly observed in the frequency histograms related to datasets with 250, 500 and 1000 points of Fig.~\ref{fig:conv_pts}. Increasing further the number of points in the dataset leads to an improved convergence rate and the error becomes close to that of the other two approaches. A similar phenomenon occurs for the global minimization approach, but in this case the snap-back branch is less pronounced and leads to smaller errors. A clustering of the results is still present in this case already from the 50-points dataset although with much lower frequencies. Moreover, the convergence rate for this approach degrades for very dense datasets (i.~e.,with more than 1000 points). The range of the solutions in Fig.~\ref{fig:conv_pts} confirms that the local minimization approach based on the consistency condition systematically overestimates the crack length, while the other two methods alternate over- and under-estimations. Also, we can see that the range of the solutions for 5000 points is indistinguishable from the reference solution for all the approaches.

\begin{figure}[!hb]
	\begin{adjustwidth}{-3cm}{-3cm}
	\centering
		\includegraphics{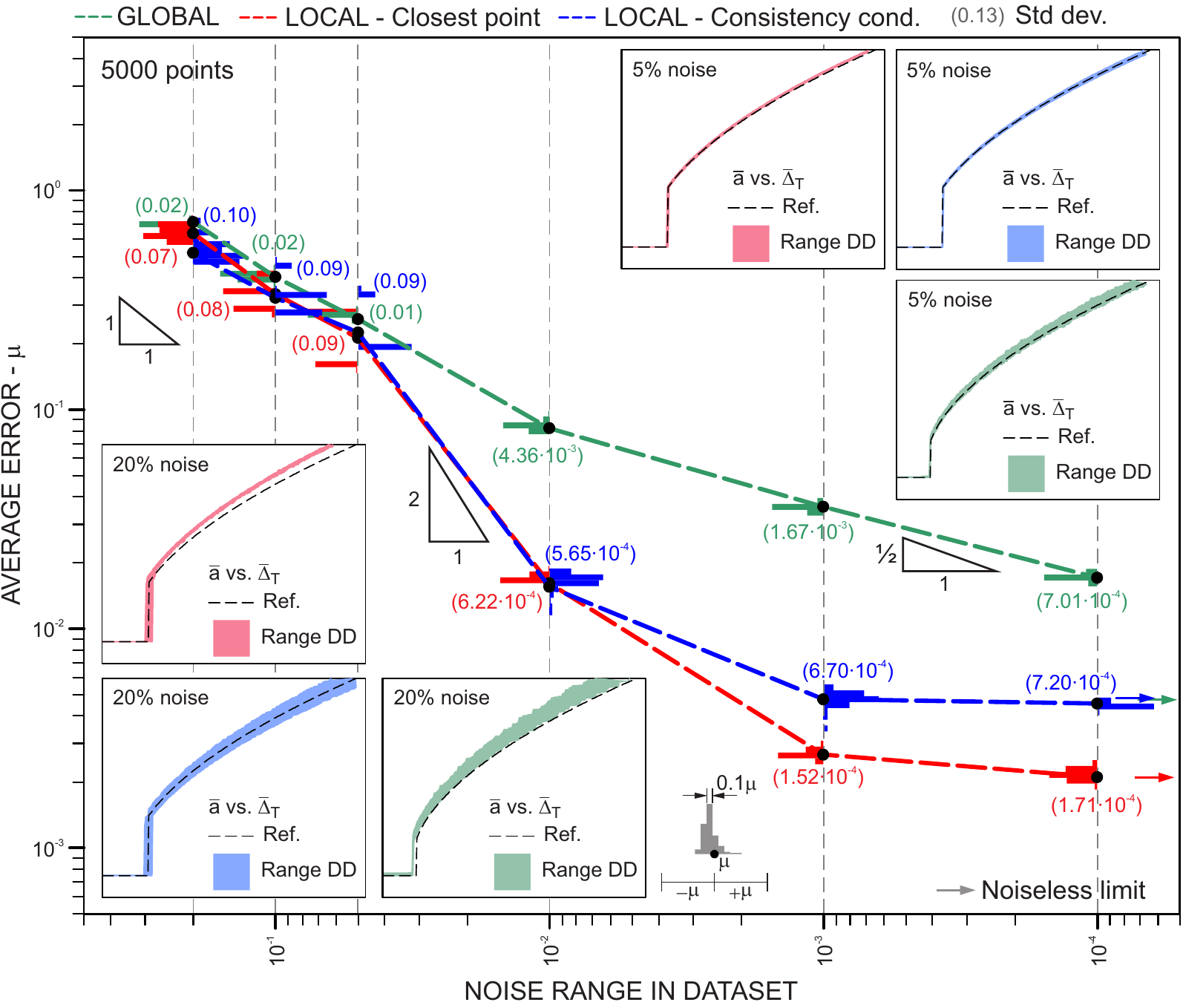}
	\end{adjustwidth}
		\caption{Convergence with respect to the noise amplitude in the dataset including frequency histograms of the error and range of the data-driven solutions for 20\% (i.~e.,$\pm$10\%) and 5\% (i.~e.,$\pm$ 2.5\%) noise amplitude.}
		\label{fig:conv_noise}
	\end{figure}

The results of the convergence study with respect to the noise amplitude are presented in Fig.~\ref{fig:conv_noise} for a dataset with 5000 points and 20\%, 10\%, 5\%, 1\%, 10$^{-1}$\% and 10$^{-2}$\% (i.~e.,$\pm$10\%, $\pm$5\%, $\pm$2.5\%, $\pm$0.5\%, $\pm$5$\cdot$10$^{-2}$\% and $\pm$5$\cdot$10$^{-3}$\%) noise ranges. For the closest point projection approach, the convergence behavior is linear up to a noise range of 5\% and then becomes almost quadratic until the error reaches the value of the noiseless case. The convergence behavior of the consistency condition approach is very similar. Unlike in the noiseless case, in presence of noise also an underestimation of the crack length is possible. Compared to both local minimization approaches, the global minimization method features a much slower convergence. This is clearly confirmed by, e.g., the data-driven solution range for 5\% noise, which is much wider than those coming from the local minimization approaches (Fig.~\ref{fig:conv_noise}). The noiseless limit is not approached even for a noise range of 10$^{-2}$\%. This slower convergence stems from the tendency of this approach, already noted earlier, to select points with high negative noise and, hence, to systematically anticipate the selection of a certain crack length (Fig.~\ref{fig:conv_noise}). For this reason, the range of the data-driven solutions lies always above the reference solution (Fig.~\ref{fig:conv_noise}).

A final comment is devoted to the computational efficiency. Although this is not quantitatively shown here, the closest point projection method is the most efficient. The other two approaches are slower due to the need of selecting points within the material data set which fulfill certain requirements. Note that in our implementation of all methods the material data set is not sorted with respect to the crack length nor efficient search algorithms are adopted. This reduces the computational speed but is consistent with the intention of the model-free data-driven philosophy not to manipulate the material data set in any way.

\FloatBarrier

\section{Summary and concluding remarks} \label{sct:conclusions}

We have proposed a data-driven approach to rate-independent fracture mechanics in brittle materials. The main idea is to remove fracture-related material modeling assumptions from the formulation and let the fracture constitutive behavior be encoded exclusively in a discrete material data set, while simultaneously keeping the epistemic laws of fracture that stem from variational principles. Here we consider both solutions based on a metastability or local stability principle, fulfilling Kuhn-Tucker conditions for the energy release rate, and solutions based on a global stability principle corresponding to the minimization of the total free energy. The data-driven solution at a given load step is identified as the point within the data set that best fulfills either stationarity or global minimization of the free energy, leading to data-driven counterparts of both approaches. For local minimization, two alternative data-driven approaches are devised, one based on the closest-point projection of the material data set onto the (analytically known) energy-release rate function and another based on a consistency condition among data points that {\sl a priori} satisfy the other two Kuhn-Tucker conditions. Both approaches have been tested on double-cantilever-beam examples with different geometries, using artificially generated material data sets, with or without random noise, which reproduce or randomize Griffith and R-curve type fracture models. A convergence study with respect of the number of points and the noise amplitude of the data set is also performed.

Based on the obtained results the following conclusions can be drawn:

\begin{itemize}
\item[-] data-driven fracture mechanics approaches based on local or global stability deliver results in excellent agreement with those of their standard fracture mechanics counterparts. This implies in particular that the known drawbacks of global minimization (most notably, the possibility for the system to overcome energy barriers in case of crack jumps) are also observed in data-driven global minimization;

\item[-] the data-driven approaches based on local minimality feature an excellent robustness with respect to noisy data, whereas the quality of data-driven global minimization results is quite sensitive to noise;

\item[-] the two devised data-driven procedures based on local minimality deliver very similar results. The procedure based on closest point projection has the advantage of being able to automatically discriminate between crack arrest and crack propagation conditions, and is computationally less expensive;

\item[-] data-driven fracture mechanics is able to correctly reproduce crack jumps with no need for {\sl ad hoc} criteria, provided that no competition takes place between different possible meta-stable states. Otherwise, the approach tends to select the state corresponding to the maximum dissipation;

\item[-] all proposed approaches deliver convergent results with respect to both the number of points in the data sets and the amplitude of a random noise. The procedure based on local minimality and closest point projection features the best convergence rate.

\end{itemize}

\FloatBarrier
\section*{Acknowledgements}
P. Carrara gratefully acknowledges the financial support of the German Research Foundation (DFG) through the Fellowship Grant CA~2359/1.

\bibliographystyle{unsrt}
\bibliography{Biblio}

\appendix

\section{Compliance for the tapered DCB} \label{app:taper_comp}

Each arm of the tapered DCB specimen of Fig.~\ref{fig:tapered} is assumed to behave as a cantilever beam whose length is equal to the crack extension $a$ and loaded with a concentrated force $P$ at the free end. The opening $\Delta$ of the tapered DCB is twice the free end deflection of the beam $v_0$, so the compliance is directly computed from (\ref{compl}) as

\eqn{eq:TDCB_compl}{C(a) = \frac{\Delta}{P}=\frac{2v_0}{P}\,.}

Neglecting the shear contribution and the deformability of the uncracked portion of the specimen, the vertical displacement $v$ can be computed integrating the Euler-Bernoulli beam equation

\eqn{eq:EB_eq}{v''(x) = \frac{M(x)}{YI(x)}=\frac{12Px}{Ybh(x)^3}\quad\text{with}\quad I(x)=\frac{bh(x)^3}{12}\,,}

\noindent where $M(x)$ is the bending moment and $I(x)$ is the moment of inertia of the rectangular DCB section.
With reference to Fig.~\ref{fig:tapered} the following quantities are defined

\eqn{eq:taper_DCB_par}{p=h_1 - m L_1\,,\quad L_{1-2}=L_1+L_T\,,\quad h_2=h_1+mL_T\,,}

\noindent while the height reads

\eqn{eq:tdcb_height}{h(x)=\begin{cases} h_1 &\text{if } 0< x \le L_1\\
									    p+mx & \text{if } L_1< x \le L_{1-2}\\
									    h_2 & \text{if } x >L_{1-2}\,.  \end{cases}}

To determine the integration constants, different kinematic boundary and continuity conditions are imposed as a function of the crack tip position, see Tab.~\ref{tab:taper_bc}.

\begin{table}[!b]
\centering
\begin{tabular}{cccc}
\toprule
 $0< a \le L_1$ & $v_1(a)=v'_1(a)=0$ &  & \\ [10pt]
$L_1< a \le L_{1-2}$ & $\begin{array}{cc}v_1(L_1)=v_T(L_1)\\v'_1(L_1)=v'_T(L_1)\end{array}$ & $v_T(a)=v'_T(a)=0$\\[10pt]
 $a >L_{1-2}$ & $\begin{array}{cc}v_1(L_1)=v_T(L_1)\\v'_1(L_1)=v'_T(L_1)\end{array}$ & $\begin{array}{cc}v_T(L_{1-2})=v_2(L_{1-2})\\v'_T(L_{1-2})=v'_2(L_{1-2})\end{array}$ &  $v_2(a)=v'_2(a)=0$  \\[10pt]
\toprule
\end{tabular}
\caption{Boundary conditions to determine the deflection of the tapered DCB specimen. Subscripts $1$, $2$ and $T$ refer to the relationships obtained integrating (\ref{eq:EB_eq}) along $L_1$, $L_2$ and $L_T$, respectively.}
\label{tab:taper_bc}
\end{table}

The compliance function is given by

\eqn{eq:taper_compl}{C(a) = \begin{cases}\displaystyle \frac{8a^3}{Ybh_1^3} &\text{if } 0< a \le L_1\\ \\
										\begin{split}\displaystyle\frac{12}{Ybm^3}\left[ \ln\left(\frac{ma+p}{mL_1+p}\right)^2 + \frac{p^2-2m^2a^2}{(ma+p)^2} +\right.\\\left.- \frac{p^2-2m^2L_1^2}{(mL_1+p)^2} + \frac{2m^3L_1^3}{3h_1^3}\right] \end{split}& \text{if } L_1< a \le L_{1-2}\\ \\
									    \begin{split}\displaystyle\frac{12}{Ybm^3}\left[ \ln\left(\frac{mL_{1-2}+p}{mL_1+p}\right)^2 + \frac{p^2-2m^2L_{1-2}^2}{(mL_{1-2}+p)^2} +\right.\\ \left.- \frac{p^2-2m^2L_1^2}{(mL_1+p)^2} + \frac{2m^3L_1^3}{3h_1^3}+ \frac{2m^3(a^3-L_{1-2}^3)}{3h_2^3}\right] \end{split} & \text{if } a >L_{1-2}\,.  \end{cases}}

\end{document}